\documentclass[journal,10pt]{IEEEtran}
\usepackage[cmex10]{amsmath}
\usepackage{graphicx}
\usepackage{algorithm}
\usepackage{algorithmic}
\usepackage{stfloats}
\usepackage{epstopdf}
\usepackage[numbers,sort&compress]{natbib}
\usepackage{multirow}
\usepackage{color}
\usepackage{balance}
\usepackage{amsmath, amssymb}
\usepackage{caption}
\usepackage{subfigure}
\usepackage{bm}
\ifCLASSINFOpdf
\else
\fi
\begin{document}
\title{\huge Weighted Sum Rate Enhancement by Using Dual-Side IOS-Assisted Full-Duplex for Multi-User MIMO Systems}
\author{Sisai~Fang,~\IEEEmembership{Student Member,~IEEE,}
Gaojie~Chen,~\IEEEmembership{Senior Member,~IEEE},
Chong~Huang,~\IEEEmembership{Member,~IEEE,}
Yue~Gao,~\IEEEmembership{Fellow,~IEEE}, Yonghui Li,~\IEEEmembership{Fellow,~IEEE}, Kai-Kit~Wong,~\IEEEmembership{Fellow,~IEEE,} \\and Jonathon A. Chambers,~\IEEEmembership{Fellow,~IEEE}
\thanks{\noindent Sisai Fang and Jonathon A. Chambers are with the School of Engineering, University of Leicester, Leicester LE1 7RH, U.K. (e-mail: sf305@leicester.ac.uk and jonathon.chambers@leicester.ac.uk.)}
\thanks{Gaojie Chen, Chong Huang and Yue Gao are with 5GIC \& 6GIC, Institute for
Communication Systems (ICS), University of Surrey, Guildford, GU2 7XH,
United Kingdom. (email: gaojie.chen@surrey.ac.uk, chong.huang@surrey.ac.uk and yue.gao@surrey.ac.uk.)}
\thanks{Yonghui Li is with the School of Electrical and Information Engineering, University of Sydney, Sydney, NSW 2006, Australia. (e-mail: yonghui.li@sydney.edu.au.)}
\thanks{Kai-Kit Wong is with the Department of Electronic and Electrical Engineering, University College London, London WC1E 7JE, U.K. (e-mail:
kai-kit.wong@ucl.ac.uk.)}
}
\maketitle

\begin{abstract}
This paper established a novel multi-input multi-output (MIMO) communication network, in the presence of full-duplex (FD) transmitters and receivers with the assistance of dual-side intelligent omni surface (IOS). Compared with the traditional IOS, the dual-side IOS allows signals from both sides to reflect and refract simultaneously, which further exploits the potential of metasurfaces to avoid frequency dependence, and size, weight, and power (SWaP) limitations. By considering both the downlink and uplink transmissions, we aim to maximize the weighted sum rate, subject to the transmit power constraints of the transmitter, the users and the dual-side reflecting and refracting phase shifts constraints. However, the formulated sum rate maximization problem is not convex, hence we exploit the weighted minimum mean square error (WMMSE) approach, and tackle the original problem iteratively by solving two sub-problems. For the beamforming matrices optimization of the downlink and uplink, we resort to the Lagrangian dual method combined with a bisection search to obtain the results. Furthermore, we resort to the quadratically constrained quadratic programming (QCQP) method to optimize the reflecting and refracting phase shifts of both sides of the IOS. Simulation results validate the efficacy of the proposed algorithm and demonstrate the superiority of the dual-side IOS.

\end{abstract}


\IEEEpeerreviewmaketitle

\section{Introduction}
As wireless communication evolves, the Internet of Things (IoT) plays a critical role in beyond fifth-generation(B5G), sixth-generation (6G), and subsequent future wireless networks, thus requiring higher speeds, connectivity, and innovative features that surpass current 5G networks to support the ever-growing traffic demands \cite{IoTJsurvey21}. Full-duplex (FD) communication shows superiority in enhancing spectrum efficiency for IoT networks, compared to conventional half-duplex (HD) communication; therefore, it has attracted tremendous attention from researchers \cite{9502651,9933358,covert}. By allowing devices to transmit and receive signals simultaneously, FD communication can ideally achieve doubled data rates. The advantages of FD communication can be concluded as 1) reducing delay, 2) improving spectrum efficiency and 3) enhancing security. Furthermore, the authors of \cite{delay} demonstrated that the end-to-end delay could be reduced using the bidirectional FD transmission. Moreover, the authors of \cite{gaojiepls} studied the physical layer security metrics of relay networks by introducing an FD jamming relay and demonstrated the secrecy rate was significantly enhanced while guaranteeing the same data rate as the HD counterpart. Also, by optimizing the beamforming design in a millimeter wave FD communication system, the rate performance was improved and the receiver-side saturation was analyzed \cite{9456023}. A deep Q-network was utilized by the authors in \cite{BufferAided} to solve problems in terms of the capacity and secrecy for cognitive radio relay networks. However, the mechanism of FD communication will inevitably introduce self-interference (SI), which is FD communication's main obstacle and limitation. As a downside of FD communication, the SI would significantly degrade the system performance if left unsolved. As claimed in \cite{self-interference}, excessive SI could reduce the capacity of FD networks to an even lower level compared to that in HD networks.

To this end, some SI cancellation (SIC) schemes are being developed to circumvent this problem, which typically is categorized as: passive, analogue and digital cancellation.  By increasing the transmit-receive antenna distance, the passive SIC can mitigate the SI because of the propagation loss, which is suitable for large-size equipment with few-antenna systems \cite{Applications}. Secondly, the analogue SIC mainly focuses on exploring the diversity bestowed by antennas, such as designing the beamforming for wireless communication systems \cite{10065530,BeamBased}. 
Thirdly, some digital SIC protocols, such as Zigzag, were utilized to suppress SI, however, the ability to cancel SI is limited due to their nonlinear SI components \cite{10258345}. It is essential to mention that traditional radio frequency (RF) SIC schemes depend on the frequency due to analogue SIC circuits, rendering the best results at the centre frequency \cite{R1}, however other results apart from the centre frequency are not promising. 
Such features in traditional SIC schemes bring severe bandwidth challenges and constraints for FD networks, which might not succeed in facilitating the operation in high-frequency bands such as millimetre waves.
The implementation and significant loss of RF SIC circuits are challenges for FD communication, and the authors showed that the operational bandwidth was restricted by hardware \cite{wideband}. 
Besides, some specific requirements in wireless communications such as the size, weight, and power (SWaP) limitations still bring challenges for conventional FD transmission, e.g. not enough RF units within limited-size equipment \cite{Non_Terrestrial} and power consumption for the SIC modules.
Overall, developing effective SIC schemes is a critical challenge for FD  wireless communications to achieve ideal data rates. 

On the other hand, the reconfigurable intelligent surface (RIS) constructed by a number of PIN diodes with the ability to guide the scattering waves has attracted tremendous attention from both industries and academia due to its energy-green, high adjustability, high portability and performance \cite{opptu}. Furthermore, the size of RIS is small; thus, it can be placed near transmitters, receivers or the surfaces of buildings. Last but not least, it was demonstrated in many works that RIS could enhance the spectrum performance of IoT networks \cite{9847234,10045778,10577673}. More recently, intelligent omni surface (IOS), also known as simultaneously transmitting and reflecting RIS (STAR-RIS) \cite{9690478}, has been developed to explore the potential of metasurfaces further \cite{9895224}. Unlike conventional RIS, the IOS is capable of reflecting and refracting signals from transmitters at the same time. In 2020, the first successful experiment of a transparent dynamic metasurface which reflects and refracts signals simultaneously at 28 GHz was implemented \cite{NTT}. Also, the authors of \cite{10370741} investigated the beamforming design and energy harvesting mechanism to improve the network data rate. Moreover, the IOS technology has been applied in secure communication networks \cite{SisaiCL}, edge computing systems \cite{9712442} and non-orthogonal multiple access (NOMA) systems \cite{10381741} to verify it is more competitive than the RIS.

To fully explore the potential of IOS and RIS in FD systems, numerous recent studies have utilized RIS or IOS to enhance the system performance in FD networks \cite{Unfolding,Aided,Achieve,Robust,9963583,twc}. 
For example, the authors of \cite{Unfolding} applied an RIS in a multi-input single-output (MISO) communication system in the presence of an FD transmitter,
and formulated a joint downlink and uplink data rate maximization problem.
In particular, the transmit power costs of both uplink and downlink transmissions were significantly reduced by introducing RISs and optimizing the transmit powers and phase shifts \cite{Aided}. 
More interestingly, the authors in \cite{Achieve} analyzed the outage and error probabilities in an FD wireless network with an RIS, and showed that RIS can effectively improve the SINR at FD nodes. 
From the security perspective, \cite{Robust} considered a practical RIS-assisted FD communication network without perfect knowledge of channel information and maximized the worst-case achievable security rate. 
Moreover, RIS was utilized to minimize the total transmission time of FD wireless-powered communication networks in IoT systems \cite{9963583}.
It is noted that the aforementioned works aimed to improve signal strength at receivers by introducing RISs, not to balance the SIC and data transmissions.
Also, it is assumed that the SI is eliminated using conventional SIC schemes in the above works, instead of suppressing it with the help of RISs.
That is, conventional SIC schemes have not solved the frequency-selectivity and SWaP problems.
To address this issue, the authors of \cite{twc} firstly applied an IOS in mitigating  SIC in FD MISO systems, however, the weighted sum rate of both uplink and downlink transmissions was not optimized, multiple receivers were not considered and the implementation required the transmitter to be on a single side of the IOS.
Furthermore, an RIS was introduced for multi-user FD systems to perform SIC \cite{WqqFD}, however, multiple antennas at the user equipmment was not investigated. 
In addition, the dual-sided IOS developed by \cite{dualside} enables signal reflection and refraction on both sides of the IOS. However, existing studies have not fully exploited its potential, particularly in FD systems, where no work has yet considered dual-sided IOS.

Motivated by the aforementioned research, we propose the design of a next-generation advanced transceiver. Specifically, we firstly introduce a dual-side IOS assisted transmitter to enhance multi-user FD multi-input multi-output (MIMO) systems, thereby boosting the data rate as well as suppressing SI for FD communication networks.
Different from the traditional IOS, dual-side IOS enables signal transmissions from both sides of the IOS rather than one.
In the meanwhile, the bandwidth and SWaP limitations that existed in conventional SIC schemes can be solved by the proposed model. 
To the best of our knowledge, we are the first to establish FD multi-user multi-input multi-output (MIMO) networks with the help of dual-side IOS to cancel SI and enhance the weighted sum rate. Specifically, we aim to maximize the joint uplink and downlink transmission data rates, subject to the power thresholds of the transmitter and users, and the amplitude, phase shift constraints for reflection and refraction of the dual-side IOS. The main contributions of this paper are listed:  
\vspace{-0.1em}
\begin{enumerate}
\item 
We establish a novel dual-side assisted FD transmitter for multi-user MIMO communication networks, where signals from the dual sides of the IOS are reflected and refracted.
The proposed scheme improves the weighted sum rate, cancels the SI as well as mitigates the SWaP limitations, meanwhile overcoming the limited ability to cancel self-interference that widely appears in the conventional SIC techniques.
\item Our main objective is to maximize the weighted sum rate by considering both the uplink and downlink transmissions, via solving the beamforming matrices optimization problem of the transmitter and the multiple users, and optimizing the phase shifts for reflection and refraction of the dual-side IOS, with transmit power and unity-modulus constraints.
\item We propose an iterative optimization method to facilitate the weighted sum rate maximization problem that is difficult to solve since the variables are coupled. Specifically, we employ the Lagrangian dual method for the beamforming matrices optimization at the transmitter and users and we use the quadratically constrained quadratic programming (QCQP) method to optimize the dual-side reflecting and refracting phase shifts optimization at the IOS.
\item The provided results demonstrate the efficacy of our proposed algorithm, and the flexibility and improvement of utilizing the dual-side IOS for FD MIMO communication networks are validated. More specifically, the weighted sum rate is enhanced significantly via introducing the dual-side IOS compared to the conventional single-side IOS.
\end{enumerate}

The organization of this paper is listed as follows. Section II shows our proposed dual-side IOS-assisted multi-user MIMO FD communication network, alongside the problem formulation.
Section III shows the proposed algorithm and solves the formulated problem iteratively, by optimizing beamforming matrices, the amplitudes, phase shifts for reflection and refraction of the dual-side IOS.
Numerical results are provided in Section IV, and demonstrate the performance improvement via introducing a dual-side IOS into an FD multi-user MIMO communication network.
In Section V, the paper is concluded.

\textit{Notations:} In this paper, scalars are denoted by italic letters, and vectors and matrices are denoted by bold lowercase letters and bold uppercase letters, respectively. $\mathbf{a}^H$ gives the Hermitian transpose of the vector $\mathbf{a}$, $\mathbf{a}^T$ and $^ - {\mathbf{a}}$ denote its transpose and conjugate operators, respectively. $\left[\mathbf{A} \right]_{i, j}$ is the element located on the $i$th row and $j$th column of matrix $\mathbf{A}$, ${\left[ {\mathbf{a}} \right]_i}$ is the $i$th element of vector ${\mathbf{a}}$. ${\mathbf{B}} \odot {{\mathbf{C}}^T}$ is the Hadamard product of ${\mathbf{B}}$ and ${\mathbf{C}}$. The trace of a matrix $\mathbf{A}$ is represented by $\text{Tr}(\mathbf{A})$. $\operatorname{Re} \left\{  \cdot  \right\}$ denotes the real part of a complex value, ${\left\| {\mathbf{A}} \right\|}$ represents the Euclidean norm of matrix $\mathbf{A}$, $\textbf{1}_M$ stands for the $M\times 1$ identity vector.  ${\text{diag}}\left\{  \cdot  \right\}$ and ${\left(  \cdot  \right)^*}$ denote the operator for diagonalization and the optimal value, respectively and $\mathcal{O}\left(  \cdot  \right)$ is the big-O notation.

\section{System Model and Problem Formulation}
\subsection{System model}
\begin{figure}[t]
        \centering
        \includegraphics*[width=80mm]{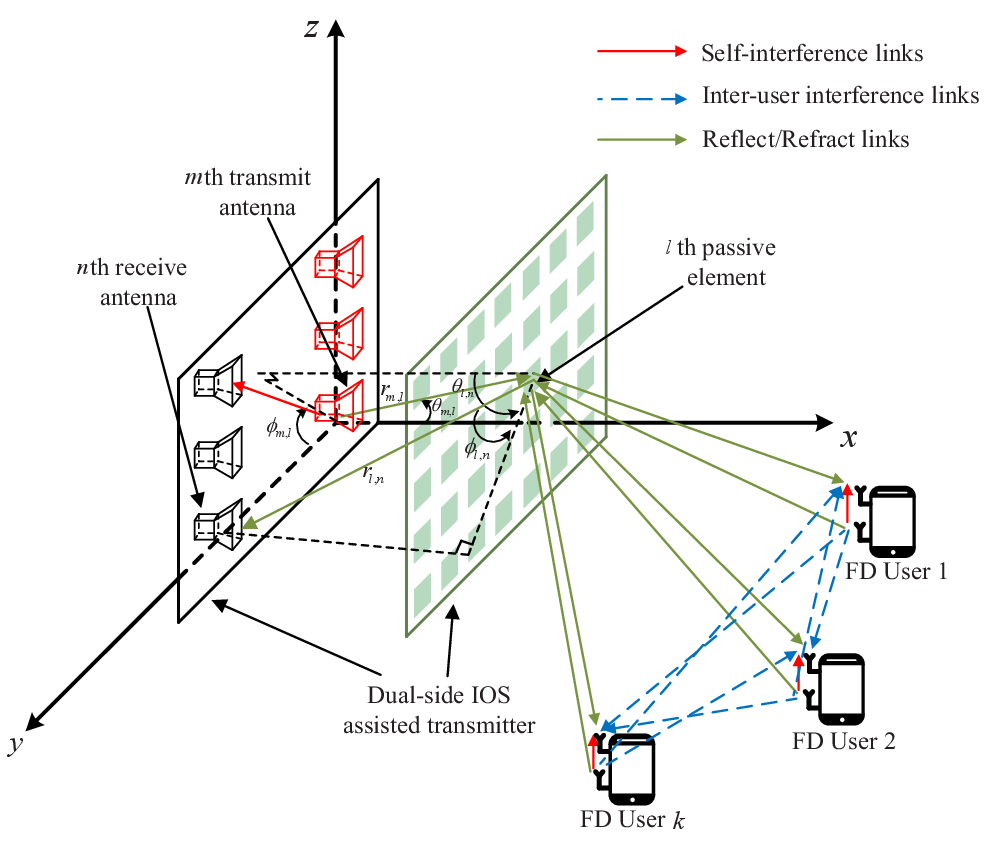}
       \caption{An illustration of a dual-side IOS-assisted FD multi-user MIMO system.}
        \label{3D}
\end{figure}

We propose  FD multi-user MIMO communication networks, with the help of a dual-side IOS as in Fig. \ref{3D}
and design the phase shifts for reflection and refraction of both sides of the IOS, to mitigate the SI as well as to boost the joint data rates of both downlink and uplink transmissions.  
It is assumed that the FD transmitter equips $N_t$ transmit antennas and $N_r$ receive antennas and is integrated with an $L$-element IOS.
In particular, each element of the dual-side IOS achieves reflection and refraction simultaneously and independently from both sides. Furthermore, the users are equipped with $N_{ut}$ transmit antennas and $N_{ur}$ receive antennas.

In addition, the phase shifts coefficient matrices for reflection and refraction of the transmitter-IOS side and the user-IOS side are represented as ${{\mathbf{\Theta }}_t} = {\text{diag}}\left\{ {{a_{t,1}}{e^{j{\alpha _{t,1}}}},{a_{t,2}}{e^{j{\alpha _{t,2}}}}, \ldots ,{a_{t,L}}{e^{j{\alpha _{t,L}}}}} \right\} \in {\mathbb{C}^{L \times L}}$, ${{\mathbf{\Phi }}_t} \!=\!{\text{diag}}\left\{ {{b_{t,1}}{e^{j{\beta _{t,1}}}},{b_{t,2}}{e^{j{\beta _{t,2}}}}, \ldots ,{b_{t,L}}{e^{j{\beta _{t,L}}}}} \right\} \!\in\! {\mathbb{C}^{L \times L}}$ and ${{\mathbf{\Theta }}_u} = {\text{diag}}\left\{ {{a_{u,1}}{e^{j{\alpha _{u,1}}}},{a_{u,2}}{e^{j{\alpha _{u,2}}}},} \right.$ $\left. { \ldots ,{a_{u,L}}{e^{j{\alpha _{u,L}}}}} \right\} \in {\mathbb{C}^{L \times L}}$, ${{\mathbf{\Phi }}_u} \!=\!{\text{diag}}\left\{ {{b_{u,1}}{e^{j{\beta _{u,1}}}},{b_{u,2}}{e^{j{\beta _{u,2}}}}, \ldots ,{b_{u,L}}{e^{j{\beta _{u,L}}}}} \right\} \!\in\! {\mathbb{C}^{L \times L}}$, respectively,
where $a_{t,l}$, $b_{t,l}$ and $a_{u,l}$, $b_{u,l}$ denote the reflecting and refracting amplitudes at the $l$th element from transmitter side and user side, respectively. $l = 1,2, \ldots , L$, and $\alpha_{t,l}$, $\beta_{t,l}$ and $\alpha_{u,l}$, $\beta_{u,l}$ presents the  $l$th element phase shift for reflection and refraction, respectively.  Based on \cite{dualside}, the constraints for the dual-side IOS can be given by
\begin{subequations}\label{constraint_1}
\begin{equation}\small\label{1a}
a_{dir,l}^2 + b_{dir,l}^2 \leq 1,
\end{equation}
\begin{equation}\small\label{1b}
0 \leq{a_{dir,l}},{b_{dir,l}} \leq 1,
\end{equation}
\begin{equation}\small\label{1c}
0 \leq {\alpha _{dir,l}},{\beta _{dir,l}} \leq 2\pi , \ \  \forall l,
\end{equation}
\end{subequations}
where $dir = \left\{ {t,u} \right\}$  denotes the transmitter side or user side, 
and the reflecting, refracting amplitudes and phase shifts are identical, hence $a_{t,l}=a_{u,l}$, $b_{t,l}=b_{u,l}$ $\alpha_{t,l}=\alpha_{u,l}$ and  $\beta_{t,l}=\beta_{u,l}$.

The downlink signals sent by the transmitter and the uplink signals sent by the $k$th user are given by
\begin{equation}\small
{{\mathbf{x}}_{{k^d}}} = {{\mathbf{V}}_{{k^d}}}{{\mathbf{s}}_{{k^d}}},
\end{equation}
\begin{equation}\small
{{\mathbf{x}}_{{k^u}}} = {{\mathbf{V}}_{{k^u}}}{{\mathbf{s}}_{{k^u}}},
\end{equation}
respectively,
where $\mathbf{V}_{k^d}\in {\mathbb{C}^{{N_t}\times {s_d}}}$ and $\mathbf{V}_{k^u}\in {\mathbb{C}^{{N_{ut}}\times {s_u}}}$ are the baseband beamforming matrices, ${s_d} = \min \left\{ {{N_t},{N_{ur}}} \right\}$ and ${s_u} = \min \left\{ {{N_r},{N_{ut}}} \right\}$ are the numbers of downlink and uplink transmission data streams, respectively.
$\mathbf{s}_{k^d}$ and $\mathbf{s}_{k^u}$ are the $k$th transmit signals at the transmitter and the user, respectively. Besides, we have $\sum\limits_{k = 1}^K {{\text{Tr}}\left( {{{\mathbf{V}}_{{k^d}}}{\mathbf{V}}_{{k^d}}^H} \right)}  \leq {P_B}$ and ${\text{Tr}}\left( {{{\mathbf{V}}_{{k^u}}}{\mathbf{V}}_{{k^u}}^H} \right) \leq {P_{{U}}}$, where ${P_B}$ and ${P_{{U}}}$ are the maximum transmit powers at the transmitter and the $k$th user.
Moreover, 
the channel gains from the transmit antennas to dual-side IOS, from the transmit to receive antennas, from the dual-side IOS to the $k$th user, from the dual-side IOS to the receive antennas, from the $j$th user to the $k$th user are denoted as ${{\mathbf{H}}_{ti}} \in {\mathbb{C}^{L \times N_{t}}}$, ${{\mathbf{H}}_{tr}} \in {\mathbb{C}^{N_{r} \times N_{t}}}$, ${{\mathbf{H}}_{iu_k}} \in {\mathbb{C}^{L \times N_{ur}}}$ and ${{\mathbf{H}}_{ir}} \in {\mathbb{C}^{L \times N_r}}$, ${{\mathbf{H}}_{{u_j}{u_k}}} \in {\mathbb{C}^{{N_{ur}} \times {N_{ut}}}}$, respectively.
Therefore, the signals received at the $k$th user and the transmitter with the IOS are denoted as
\begin{equation}\small\label{signal_d}
{{\mathbf{y}}_{{u_k}}} \!\!=\! \underbrace {{\mathbf{H}}_{i{u_k}}^H\!{{\mathbf{\Phi }}_t}\!{{\mathbf{H}}_{ti}}\!{{\mathbf{x}}_{{k^d}}}}_{{u_{k1}}} \!+\! \!\underbrace {\sum\limits_{j = 1}^K {{{\mathbf{H}}_{{u_j}{u_k}}}\!{{\mathbf{x}}_{{k^u}}}} }_{{u_{k2}}} \!+\! \underbrace {\sum\limits_{j = 1}^K {{\mathbf{H}}_{i{u_k}}^H\!{{\mathbf{\Theta }}_u}{{\mathbf{H}}_{{u_j}i}}{{\mathbf{x}}_{{k^u}}}} }_{{u_{k3}}} \!+  {{\mathbf{n}}_{{u_k}}}, 
\end{equation}
\begin{equation}\small\label{signal_r}
{{\mathbf{y}}_r} = \underbrace {\sum\limits_{j = 1}^K {{\mathbf{H}}_{ir}^H{{\mathbf{\Phi }}_u}{{\mathbf{H}}_{{u_j}i}}{{\mathbf{x}}_{{k^u}}}} }_{r_1} + \underbrace {{{\mathbf{H}}_{tr}}{{\mathbf{x}}_{{k^d}}}}_{r_2} + \underbrace {{\mathbf{H}}_{ir}^H{{\mathbf{\Theta }}_t}{{\mathbf{H}}_{ti}}{{\mathbf{x}}_{{k^d}}}}_{r_3} + {{\mathbf{n}}_r},
\end{equation}respectively, where $u_{k1}$, $u_{k2}$ and $u_{k3}$ denote the signals via the transmitter-IOS-$k$th user, the $j$th user-$k$th user, and $j$th user-IOS-$k$th user, respectively. Besides, $r_1$, $r_2$ and $r_3$ denote the signals via the users-IOS-transmitter, the transmit-receive antennas, and transmit antennas-IOS-receive antennas, respectively. It is noted that $u_{k3}$ and $r_3$ are undesired information and regarded as interference at the $k$th user and the transmitter, respectively.
Besides, ${\mathbf{n}_{u_k}} \sim \mathcal{CN}\left( {0,\sigma _{u_k}^2{\mathbf{I}}} \right)$ and ${\mathbf{n}_r} \sim \mathcal{CN}\left( {0,\sigma _r^2{\mathbf{I}}} \right)$ 
denote the additive Gaussian noise at the $k$th user and the transmitter, respectively.
The noise powers at the $k$th user and the transmitter are denoted as $\sigma _{u_k}^2$ and $\sigma _r^2$, respectively.
In addition, as shown in Fig. \ref{3D}, the positions of the transmit antennas are characterized as $\left( {{r_{l,m}},\theta _{l,m},\phi _{l,m}} \right)$, ${r_{l,m}} \geq 0$, $\theta _{l,m} \in \left[ {0,\frac{\pi }{2}} \right]$, $\phi _{l,m} \in \left[ {0,2\pi } \right]$ in $L$ different spherical coordinate systems whose origins are the positions of the dual-side IOS elements \cite{FDmainref}. Specifically, ${\theta _{l,m}}$, ${\phi _{l,m}}$ and ${r_{l,m}}$ represent the  elevation angle, azimuth angle of the $m$th transmit antenna, respectively, and ${r_{l,m}}$ is the distance between the $l$th IOS element and the $m$th transmit antenna.
Similarly,  the positions of the receive antennas at the elements of the IOS, that of the receive antennas at the transmit antennas, that of the transmit antennas at the receive antennas are denoted as  $\left( {{r_{l,n}},{\theta _{l,n}},{\phi _{l,n}}} \right)$,
$\left( {{r_{m,n}},{\theta _{m,n}},{\phi _{m,n}}} \right)$ and $\left( {{r_{n,m}},{\theta _{n,m}},{\phi _{n,m}}} \right)$, respectively,
where ${r_{l,n}},{r_{m,n}},{r_{n,m}} \geq 0$,
${\theta _{l,n}},{\theta _{m,n}},{\theta _{n,m}} \in \left[ {0,\frac{\pi }{2}} \right]$, ${\phi _{l,n}},{\phi _{m,n}},{\phi _{n,m}} \in \left[ {0,2\pi } \right]$. 
Thus the corresponding channel coefficients are defined as follows:
\begin{equation}\small
{{\mathbf{H}}_{ti}}\! =\! {\left[ {\frac{{\lambda\! \sqrt {{G^t}\left( {{\theta _{l,m}},{\phi _{l,m}}} \right)\!} }}{{4\pi {r_{l,m}}}}\!{e^{ - j\frac{{2\pi {r_{l,m}}}}{\lambda }}}} \right]}_{l,m},
\end{equation}
\begin{equation}\small
\begin{gathered}
  {{\mathbf{H}}_{tr}} = \left[ {\frac{{\lambda \sqrt {{G^t}\left( {{\theta _{m,n}},{\phi _{m,n}}} \right){G^r}\left( {{\theta _{n,m}},{\phi _{n,m}}} \right)} }}{{4\pi r_{m,n}^{{\kappa  \mathord{\left/
 {\vphantom {\kappa  2}} \right.
 \kern-\nulldelimiterspace} 2}}}}} \right. \hfill \\
 \qquad\quad {\left. { \times \left( {\sqrt {\frac{\chi }{{\chi  + 1}}} {e^{ - j\frac{{2\pi {r_{m,n}}}}{\lambda }}} + \sqrt {\frac{1}{{\chi  + 1}}} h_{tr}^{nlos}} \right)} \right]_{n,m}}, \hfill \\ 
\end{gathered} 
\end{equation}
\begin{equation}\small
{{\mathbf{H}}_{i{u_j}}} \!=\! {{\mathbf{H}}_{{u_j}i}^T}\! = \!\!\left[ {\frac{\lambda }{{4\pi r_{l,u_{j}}^{{\kappa  \mathord{\left/
 {\vphantom {\kappa  2}} \right.
 \kern-\nulldelimiterspace} 2}}}}\left( {\sqrt {\frac{\chi}{{\chi \!\!+\!\! 1}}} {e^{ - j\frac{{2\pi {r_{l,u_{j}}}}}{\lambda }}} \!+\! \sqrt {\frac{1}{{\chi \!\!+\!\! 1}}} h_{iu_{j}}^{nlos}} \right)}\!\! \right]_{u_{j},l},
\end{equation}
\begin{equation}\small
{{\mathbf{H}}_{ir}} = {\left[ {\frac{{\lambda \sqrt {{G^r}\left( {{\theta _{l,n}},{\phi _{l,n}}} \right)} }}{{4\pi {r_{l,n}}}}{e^{ - j\frac{{2\pi {r_{l,n}}}}{\lambda }}}} \right]},
\end{equation}
\begin{equation}\small
{{\mathbf{H}}_{{u_j}{u_k}}} \!=\! \left[ {\frac{\lambda }{{4\pi {r_{{u_j}{u_k}}}}}\left( {\sqrt {\frac{\chi}{{\chi + 1}}} {e^{ - j\frac{{2\pi {r_{j,k}}}}{\lambda }}} + \sqrt {\frac{1}{{\chi + 1}}} {h}_{{u_j}{u_k}}^{nlos}} \right)} \right],
\end{equation}
where $\lambda $ represents the carrier wavelength, ${G^t}\left( {\theta ,\phi } \right)$ and ${G^r}\left( {\theta ,\phi } \right)$  denote the the transmit antenna gain and receive antenna gain,
\footnote{The impact of the elevation angles on antenna gains, and the gain model are shown in \cite{FDmainref}.} respectively.
Furthermore, $\chi$ and $\kappa$ denote the Rician factor and the exponents of channel fading, respectively. $r_{l,u_{j}}$ is the distance between the $l$th element and the $j$th user,  $h_{tr}^{nlos}$, $h_{iu_{j}}^{nlos}$ and $h_{u_{j}u_{k}}^{nlos}$ represent the corresponding  non line-of-sight channel components,\footnote{
In this paper, it is assumed that the channel state information (CSI) of transmitter-IOS-user and transmit-receive antennas is accessible at the dual-side IOS-aided transmitter.  \cite{Chen,qingqingWu,CSI2} provide the details regarding the channel estimation, which is beyond the scope of this paper.} respectively.

\subsection{Problem formulation}
In this paper, our objective is  to maximize the joint uplink and downlink data rate, by obtaining the optimized beamforming matrices at the transmitter, the transmit power  at the users, and the phase shift unity-modulus constraints of the dual-side IOS. Firstly, we can obtain the downlink and uplink rates as follows:
\begin{equation}\small\label{R_ud}
R_{{u_k}}^d \!=\! {\log _2}\left| {{\mathbf{I}} \!+\! {{\mathbf{H}}_{{k^d}}}\!{{\mathbf{V}}_{{k^d}}}\!{\mathbf{V}}_{{k^d}}^H\!{\mathbf{H}}_{{k^d}}^H\!{{\left( {\sum\limits_{j = 1}^K\! {{{\mathbf{H}}_{jk}}\!{{\mathbf{V}}_{{j^u}}}\!{\mathbf{V}}_{{j^u}}^H\!{\mathbf{H}}_{jk}^H \!\!+\!\! \sigma _{{u_k}}^2{\mathbf{I}}} } \right)}^{ \!-\! 1}}} \right|,
\end{equation}
\begin{equation}\small\label{R_uu}
\begin{gathered}
  R_{{u_k}}^u = {\log _2}\left| {{\mathbf{I}} + {{\mathbf{H}}_{{k^u}}}{{\mathbf{V}}_{{k^u}}}{\mathbf{V}}_{{k^u}}^H{\mathbf{H}}_{{k^u}}^H\left( {\sum\limits_{j = 1\left( {j \ne k} \right)}^K {{{\mathbf{H}}_{{j^u}}}{{\mathbf{V}}_{{j^u}}}{\mathbf{V}}_{{j^u}}^H{\mathbf{H}}_{{j^u}}^H} } \right.} \right. \hfill \\
 \qquad\qquad\qquad   \left. {\left. { + \sum\limits_{j = 1}^K {{{\mathbf{H}}_t}{{\mathbf{V}}_{{j^d}}}{\mathbf{V}}_{{j^d}}^H{\mathbf{H}}_t^H}  + \sigma _r^2{\mathbf{I}}} \right)} \right|, \hfill \\ 
\end{gathered} 
\end{equation}
where ${{\mathbf{H}}_{{k^d}}} = {\mathbf{H}}_{i{u_k}}^H{{\mathbf{\Phi }}_t}{{\mathbf{H}}_{ti}}$,
${{\mathbf{H}}_{jk}} = {{\mathbf{H}}_{{u_j}{u_k}}} + {\mathbf{H}}_{i{u_k}}^H{{\mathbf{\Theta }}_u}{{\mathbf{H}}_{{u_j}i}}$, ${{\mathbf{H}}_{{k^u}}} = {\mathbf{H}}_{ir}^H{{\mathbf{\Phi }}_u}{{\mathbf{H}}_{{u_k}i}}$ and ${{\mathbf{H}}_t} = {{\mathbf{H}}_{tr}} + {\mathbf{H}}_{ir}^H{{\mathbf{\Theta }}_t}{{\mathbf{H}}_{ti}}$. From the aforementioned analysis, the formulated problem is given as follows:

\vspace{-0.4em}
\begin{small}\begin{align}
\mathop {\max }\limits_{\mathbf{\Theta}_t,\mathbf{\Phi}_t,\mathbf{\Theta}_u,\mathbf{\Phi}_u,\mathbf{V}_{k^d},\mathbf{V}_{k^u}}&\ \ {\sum\limits_{k = 1}^K {\gamma _{k^d}} R_{{u_k}}^d + \gamma _{k^u}R_{{u_k}}^u}  \label{P1_OF}\\
{\text{s.t.}}\ \ & \ \   \sum\limits_{k = 1}^K {{\text{Tr}}\left( {{{\mathbf{V}}_{{k^d}}}{\mathbf{V}}_{{k^d}}^H} \right) \leq {P_B}} ,  \tag{\ref{P1_OF}{a}}  \label{P1_Power} \\
& \ \  {\text{Tr}}\left( {{{\mathbf{V}}_{{k^u}}}{\mathbf{V}}_{{k^u}}^H} \right) \leq {P_{{U}}}, \quad \forall k,  \tag{\ref{P1_OF}{b}}  \label{P1_sinr}\\
& \ \  a_{dir,l}^2 + b_{dir,l}^2 \leq 1, \tag{\ref{P1_OF}{c}}  \label{P1_ampli} \\
& \ \  0 \leq{a_{dir,l}},{b_{dir,l}} \leq 1, \tag{\ref{P1_OF}{d}}  \label{P1_ampli2}\\
& \ \  0 \leq {\alpha _{dir,l}},{\beta _{dir,l}} \leq 2\pi , \ \  \forall l,  \tag{\ref{P1_OF}{e}}  \label{P1_phase}
\end{align}\end{small}and ${\gamma _{{k^d}}},{\gamma _{{k^u}}} \in \left( {0,1} \right)$ are the weighted factors for the $k$th downlink and uplink transmissions based on the requirement of QoS in different scenarios, respectively.

\section{Optimization for the Weighted Sum Rate Enhancement}
This section aims to maximize the joint data rates of downlink and uplink transmissions, which is the weighted sum rate, by optimizing the phase shifts of the dual-side IOS $\mathbf{\Theta}_t$, $\mathbf{\Theta}_u$, $\mathbf{\Phi}_t$, $\mathbf{\Phi}_u$, and the beam steering matrices ${{\mathbf{V}}_{k^d}}$ and ${{\mathbf{V}}_{k^u}}$ for the downlink and uplink transmissions iteratively.
\subsection{Problem Reformulation}
Firstly, by using the first-order optimality condition, exploiting the matrix feature between the sum rate and the mean-square error (MSE) for the optimal decoding matrices, we can reformulate the $k$th downlink rate function.
Specifically, with the linear decoding matrix ${{\mathbf{U}}_{k^d}} \in {\mathbb{C}^{{N_t} \times s_d}}$, the MSE of estimation \cite{UcWc} is expressed as 
\begin{equation}\small\label{E_kd}
\begin{gathered}
  {{\mathbf{E}}_{{k^d}}} = \left( {{\mathbf{U}}_{{k^d}}^H{{\mathbf{H}}_{{k^d}}}{{\mathbf{V}}_{{k^d}}} - {\mathbf{I}}} \right){\left( {{\mathbf{U}}_{{k^d}}^H{{\mathbf{H}}_{{k^d}}}{{\mathbf{V}}_{{k^d}}} - {\mathbf{I}}} \right)^H} \hfill \\
  \qquad\qquad+ {\mathbf{U}}_{{k^d}}^H\left( {\sum\limits_{j = 1}^K {{{\mathbf{H}}_{jk}}{{\mathbf{V}}_{{j^u}}}{\mathbf{V}}_{{j^u}}^H{\mathbf{H}}_{jk}^H + \sigma _{{u_k}}^2{\mathbf{I}}} } \right){{\mathbf{U}}_{{k^d}}}. \hfill \\ 
\end{gathered} 
\end{equation}
By introducing an auxiliary variable ${\mathbf{W}_{k^d}} \succeq 0$, ${{\mathbf{W}}_{{k^d}}} \in {\mathbb{C}^{{s_d} \times {s_d}}}$, the rate of the $k$th downlink transmission is denoted by
\begin{equation}\small\label{f_kd}
\begin{gathered}
  h_{{u^k}}^d\left( {{{\mathbf{W}}_{{k^d}}},{{\mathbf{U}}_{{k^d}}},{{\mathbf{V}}_{{k^d}}},{{\mathbf{V}}_{{k^u}}}} \right) \hfill \\
   \triangleq \mathop {\max }\limits_{{{\mathbf{W}}_{{k^d}}},{{\mathbf{U}}_{{k^d}}}} \log \left| {{{\mathbf{W}}_{{k^d}}}} \right| - {\text{Tr}}\left( {{{\mathbf{W}}_{{k^d}}}{{\mathbf{E}}_{{k^d}}}\left( {{{\mathbf{U}}_{{k^d}}},{{\mathbf{V}}_{{k^d}}},{{\mathbf{V}}_{{k^u}}}} \right)} \right) + {s_d}, \hfill \\ 
\end{gathered} 
\end{equation}
and the optimal ${\mathbf{U}}_{k^d}^*$ and $\mathbf{W}_{k^d}^*$ for the maximum value (similar to \cite{UcWc}) of $h_{{u^k}}^d\!\left( {{{\mathbf{W}}_{{k^d}}},\!{{\mathbf{U}}_{{k^d}}},\!{{\mathbf{V}}_{{k^d}}},\!{{\mathbf{V}}_{{k^u}}}} \right)$ are given by
\begin{equation}\small\label{U_kd}
{{\mathbf{U}}_{k^d}^*} \!=\!\! {\left( \!{{{\mathbf{H}}_{{k^d}}}\!{{\mathbf{V}}_{{k^d}}}\!{\mathbf{V}}_{{k^d}}^H\!{\mathbf{H}}_{{k^d}}^H \!+\! \sum\limits_{j = 1}^K {{{\mathbf{H}}_{jk}}\!{{\mathbf{V}}_{{j^u}}}{\mathbf{V}}_{{j^u}}^H\!{\mathbf{H}}_{jk}^H \!+\! \sigma _{{u_k}}^2{\mathbf{I}}\!} } \right)^{ \!- 1}}\!{{\mathbf{H}}_{{k^d}}}\!{{\mathbf{V}}_{{k^d}}},
\end{equation}
\begin{equation}\small\label{W_kd}
\begin{gathered}
  {\mathbf{W}}_{{k^d}}^* = \left( {\left( {{\mathbf{U}}_{{k^d}}^{*H}{{\mathbf{H}}_{{k^d}}}{{\mathbf{V}}_{{k^d}}} - {\mathbf{I}}} \right){{\left( {{\mathbf{U}}_{{k^d}}^{*H}{{\mathbf{H}}_{{k^d}}}{{\mathbf{V}}_{{k^d}}} - {\mathbf{I}}} \right)}^H}} \right. \hfill \\
 \qquad\qquad {\left. { + {\mathbf{U}}_{{k^d}}^{*H}\left( {\sum\limits_{j = 1}^K {{{\mathbf{H}}_{jk}}{{\mathbf{V}}_{{j^u}}}{\mathbf{V}}_{{j^u}}^H{\mathbf{H}}_{jk}^H + \sigma _{{u_k}}^2{\mathbf{I}}} } \right){\mathbf{U}}_{{k^d}}^*} \right)^{ - 1}}, \hfill \\ 
\end{gathered} 
\end{equation}
By substituting \eqref{E_kd} into \eqref{f_kd}, we obtain the rate of the $k$th downlink transmission as follows:
\begin{equation}\small\label{h_ukd}
\begin{gathered}
  h_{{u^k}}^d\left( {{{\mathbf{W}}_{{k^d}}},{{\mathbf{U}}_{{k^d}}},{{\mathbf{V}}_{{k^d}}},{{\mathbf{V}}_{{k^u}}}} \right) \hfill \\
   \triangleq \log \left| {{{\mathbf{W}}_{{k^d}}}} \right| - {\text{Tr}}\left( {{{\mathbf{W}}_{{k^d}}}{\mathbf{U}}_{{k^d}}^H{{\mathbf{H}}_{{k^d}}}{{\mathbf{V}}_{{k^d}}}{\mathbf{V}}_{{k^d}}^H{\mathbf{H}}_{{k^d}}^H{{\mathbf{U}}_{{k^d}}}} \right) \hfill \\
 \quad  + {\text{Tr}}\left( {{{\mathbf{W}}_{{k^d}}}{\mathbf{U}}_{{k^d}}^H{{\mathbf{H}}_{{k^d}}}{{\mathbf{V}}_{{k^d}}}} \right){\text{ + Tr}}\left( {{{\mathbf{W}}_{{k^d}}}{\mathbf{V}}_{{k^d}}^H{\mathbf{H}}_{{k^d}}^H{{\mathbf{U}}_{{k^d}}}} \right) \hfill \\
 \quad  - \sum\limits_{j = 1}^K {{\text{Tr}}\left( {{{\mathbf{W}}_{{k^d}}}{\mathbf{U}}_{{k^d}}^H{{\mathbf{H}}_{jk}}{{\mathbf{V}}_{{j^u}}}{\mathbf{V}}_{{j^u}}^H{\mathbf{H}}_{jk}^H{{\mathbf{U}}_{{k^d}}}} \right)} \; \hfill \\
 \quad  - {\text{ Tr}}\left( {{{\mathbf{W}}_{{k^d}}}{\mathbf{U}}_{{k^d}}^H\sigma _{{u_k}}^2{{\mathbf{U}}_{{k^d}}}} \right) - {\text{Tr}}\left( {{{\mathbf{W}}_{{k^d}}}} \right) + {s_d}. \hfill \\ 
\end{gathered} 
\end{equation}

Similarly, by introducing ${{\mathbf{U}}_{k^u}} \in {\mathbb{C}^{{N_{ut}} \times s_u}}$ and auxiliary variables ${{\mathbf{W}}_{k^u}} \in {\mathbb{C}^{s_u \times s_u}}$, we have
\begin{equation}\small\label{g_ku}
\begin{gathered}
  h_{{u^k}}^u\left( {{{\mathbf{W}}_{{k^u}}},{{\mathbf{U}}_{{k^u}}},{{\mathbf{V}}_{{k^d}}},{{\mathbf{V}}_{{k^u}}}} \right) \hfill \\
   \triangleq \mathop {\max }\limits_{{{\mathbf{W}}_{{k^u}}},{{\mathbf{U}}_{{k^u}}}} \log \left| {{{\mathbf{W}}_{{k^u}}}} \right| - {\text{Tr}}\left( {{{\mathbf{W}}_{{k^u}}}{{\mathbf{E}}_{{k^u}}}\left( {{{\mathbf{U}}_{{k^u}}},{{\mathbf{V}}_{{k^d}}},{{\mathbf{V}}_{{k^u}}}} \right)} \right) + {s_u}. \hfill \\ 
\end{gathered} 
\end{equation}
Then we have
\begin{equation}\small\label{E_ku}
\begin{gathered}
  {{\mathbf{E}}_{{k^u}}} = \left( {{\mathbf{U}}_{{k^u}}^H{{\mathbf{H}}_{{k^u}}}{{\mathbf{V}}_{{k^u}}} - {\mathbf{I}}} \right){\left( {{\mathbf{U}}_{{k^u}}^H{{\mathbf{H}}_{{k^u}}}{{\mathbf{V}}_{{k^u}}} - {\mathbf{I}}} \right)^H} + {\mathbf{U}}_{{k^u}}^H \hfill \\
  \quad \times \left( {\sum\limits_{j = 1\left( {j \ne k} \right)}^K \! {{{\mathbf{H}}_{{j^u}}} \!{{\mathbf{V}}_{{j^u}}} \!{\mathbf{V}}_{{j^u}}^H \!{\mathbf{H}}_{{j^u}}^H \! + \! \sum\limits_{j = 1}^K {{{\mathbf{H}}_t} \!{{\mathbf{V}}_{{j^d}} \!}{\mathbf{V}}_{{j^d}}^H \!{\mathbf{H}}_t^H  \!+ \! } \sigma _r^2 \!{\mathbf{I}} \!} } \right) \!{{\mathbf{U}}_{{k^u}}}, \hfill \\ 
\end{gathered} \end{equation}
\begin{equation}\small\label{U_ku}
\begin{gathered}
  {\mathbf{U}}_{{k^u}}^* = \left( {{{\mathbf{H}}_{{k^u}}}{{\mathbf{V}}_{{k^u}}}{\mathbf{V}}_{{k^u}}^H{\mathbf{H}}_{{k^u}}^H + \sum\limits_{j = 1\left( {j \ne k} \right)}^K {{{\mathbf{H}}_{{j^u}}}{{\mathbf{V}}_{{j^u}}}{\mathbf{V}}_{{j^u}}^H{\mathbf{H}}_{{j^u}}^H} } \right. \hfill \\
 \qquad\qquad {\left. { + \sum\limits_{j = 1}^K {{{\mathbf{H}}_t}{{\mathbf{V}}_{{j^d}}}{\mathbf{V}}_{{j^d}}^H{\mathbf{H}}_t^H + } \sigma _r^2{\mathbf{I}}} \right)^{ - 1}}{{\mathbf{H}}_{{k^u}}}{{\mathbf{V}}_{{k^u}}}, \hfill \\ 
\end{gathered} 
\end{equation}
\begin{equation}\small\label{W_ku}
{\mathbf{W}}_{{k^u}}^* = {\left( \begin{gathered}
  \left( {{\mathbf{U}}_{{k^u}}^{*H}{{\mathbf{H}}_{{k^u}}}{{\mathbf{V}}_{{k^u}}} - {\mathbf{I}}} \right){\left( {{\mathbf{U}}_{{k^u}}^{*H}{{\mathbf{H}}_{{k^u}}}{{\mathbf{V}}_{{k^u}}} - {\mathbf{I}}} \right)^H} \hfill \\
   + {\mathbf{U}}_{{k^u}}^{*H}\left( {\sum\limits_{j = 1\left( {j \ne k} \right)}^K {{{\mathbf{H}}_{{j^u}}}{{\mathbf{V}}_{{j^u}}}{\mathbf{V}}_{{j^u}}^H{\mathbf{H}}_{{j^u}}^H} } \right. \hfill \\
  \left. { + \sum\limits_{j = 1}^K {{{\mathbf{H}}_t}{{\mathbf{V}}_{{j^d}}}{\mathbf{V}}_{{j^d}}^H{\mathbf{H}}_t^H + } \sigma _r^2{\mathbf{I}}} \right){\mathbf{U}}_{{k^u}}^* \hfill \\ 
\end{gathered}  \right)^{ - 1}}.
\end{equation}
By substituting \eqref{E_ku} into \eqref{g_ku}, we obtain the rate of the $k$th uplink transmission as follows:
\begin{equation}\small\label{h_uku}
\begin{gathered}
  h_{{u^k}}^u\left( {{{\mathbf{W}}_{{k^u}}},{{\mathbf{U}}_{{k^u}}},{{\mathbf{V}}_{{k^d}}},{{\mathbf{V}}_{{k^u}}}} \right) \hfill \\
   \!\triangleq \log \left| {{{\mathbf{W}}_{{k^u}}}} \right| + {\text{Tr}}\left( {{{\mathbf{W}}_{{k^u}}}{\mathbf{U}}_{{k^u}}^H{{\mathbf{H}}_{{k^u}}}{{\mathbf{V}}_{{k^u}}}} \right) \!+\! {\text{Tr}}\left(\! {{{\mathbf{W}}_{{k^u}}}\!{\mathbf{V}}_{{k^u}}^H\!{\mathbf{H}}_{{k^u}}^H{{\mathbf{U}}_{{k^u}}}}\! \right) \hfill \\
  \quad -\! \sum\limits_{j = 1}^K {{\text{Tr}}\left( \!{{{\mathbf{W}}_{{k^u}}}{\mathbf{U}}_{{k^u}}^H\!{{\mathbf{H}}_{{j^u}}}\!{{\mathbf{V}}_{{j^u}}}{\mathbf{V}}_{{j^u}}^H\!{\mathbf{H}}_{{j^u}}^H\!{{\mathbf{U}}_{{k^u}}}} \!\right)} \! -\! {\text{Tr}}\left(\! {{{\mathbf{W}}_{{k^u}}}\!{\mathbf{U}}_{{k^u}}^H\!\sigma _r^2\!{{\mathbf{U}}_{{k^u}}}} \!\right) \hfill \\
  \quad  - \sum\limits_{j = 1}^K {{\text{Tr}}\left( {{{\mathbf{W}}_{{k^u}}}{\mathbf{U}}_{{k^u}}^H{{\mathbf{H}}_t}{{\mathbf{V}}_{{j^d}}}{\mathbf{V}}_{{j^d}}^H{\mathbf{H}}_t^H{{\mathbf{U}}_{{k^u}}}} \right)} \! -\! {\text{Tr}}\left( {{{\mathbf{W}}_{{k^u}}}} \right) + {s_u}. \hfill \\ 
\end{gathered} 
\end{equation}
Therefore, by combining the downlink rate of \eqref{h_ukd} and uplink rate of \eqref{h_uku} into the joint data rates of downlink and uplink transmissions in \eqref{P1_OF},
the problem can be rewritten as

\vspace{-0.4em}
\begin{small}
\begin{align}
\mathop {\max }\limits_{\mathbf{S}}&\ \ {R_w} \label{P2_OF}\\
{\text{s.t.}}\ \ & \ \   \sum\limits_{k = 1}^K {{\text{Tr}}\left( {{{\mathbf{V}}_{{k^d}}}{\mathbf{V}}_{{k^d}}^H} \right) \leq {P_B}} ,  \tag{\ref{P2_OF}{a}}  \label{P2_Power} \\
& \ \  {\text{Tr}}\left( {{{\mathbf{V}}_{{k^u}}}{\mathbf{V}}_{{k^u}}^H} \right) \leq {P_{{U}}}, \ \   \forall k, \tag{\ref{P2_OF}{b}}  \label{P2_sinr}\\
& \ \  a_{dir,l}^2 + b_{dir,l}^2 \leq 1, \tag{\ref{P2_OF}{c}}  \label{P2_ampli} \\
& \ \  0 \leq{a_{dir,l}},{b_{dir,l}} \leq 1, \tag{\ref{P2_OF}{d}}  \label{P2_ampli2}\\
& \ \  0 \leq {\alpha _{dir,l}},{\beta _{dir,l}} \leq 2\pi , \ \  \forall l,  \tag{\ref{P2_OF}{e}}  \label{P2_phase}
\end{align}\end{small}where ${\mathbf{S}} \!= \!\left\{ {{{\mathbf{\Theta }}_t},\!{{\mathbf{\Phi }}_t},\!{{\mathbf{\Theta }}_u},\!{{\mathbf{\Phi }}_u},\!{{\mathbf{V}}_{{k^d}}},\!{{\mathbf{V}}_{{k^u}}},\!{{\mathbf{U}}_{{k^d}}},\!{{\mathbf{W}}_{{k^d}}},\!{{\mathbf{U}}_{{k^u}}},\!{{\mathbf{W}}_{{k^u}}}} \!\right\}$ and
\begin{equation}\small\label{R_w}
\begin{gathered}
  {R_w} \hfill \\
   = \sum\limits_{k = 1}^K {{\gamma _{{k^d}}}} {\text{Tr}}\left( {{{\mathbf{W}}_{{k^d}}}\!{\mathbf{V}}_{{k^d}}^H\!{\mathbf{H}}_{{k^d}}^H\!{{\mathbf{U}}_{{k^d}}}} \right) \!+\! \sum\limits_{k = 1}^K\! {{\gamma _{{k^d}}}{\text{Tr}}\left(\! {{{\mathbf{W}}_{{k^d}}}\!{\mathbf{U}}_{{k^d}}^H\!{{\mathbf{H}}_{{k^d}}}\!{{\mathbf{V}}_{{k^d}}}} \!\right)}  \hfill \\
  \quad + \!\sum\limits_{k = 1}^K \!{{\gamma _{{k^u}}}\!{\text{Tr}}\left(\! {{{\mathbf{W}}_{{k^u}}}\!{\mathbf{U}}_{{k^u}}^H\!{{\mathbf{H}}_{{k^u}}}\!{{\mathbf{V}}_{{k^u}}}}\! \right)\! +\! \sum\limits_{k \!=\! 1}^K {{\gamma _{{k^u}}}\!{\text{Tr}}\left( \!{{{\mathbf{W}}_{{k^u}}}\!{\mathbf{V}}_{{k^u}}^H\!{\mathbf{H}}_{{k^u}}^H\!{{\mathbf{U}}_{{k^u}}}}\! \right)} }  \hfill \\
  \quad  -\! \sum\limits_{k \!=\! 1}^K {{\gamma _{{k^d}}}} \!{\text{Tr}}\left(\! {{{\mathbf{W}}_{{k^d}}}\!{\mathbf{U}}_{{k^d}}^H\!{{\mathbf{H}}_{{k^d}}}\!{{\mathbf{V}}_{{k^d}}}\!{\mathbf{V}}_{{k^d}}^H\!{\mathbf{H}}_{{k^d}}^H\!{{\mathbf{U}}_{{k^d}}}} \!\right) \hfill \\
  \quad  -\! \sum\limits_{k = 1}^K {{\gamma _{{k^d}}}} \!\sum\limits_{j = 1}^K {{\text{Tr}}\left( \!{{{\mathbf{W}}_{{k^d}}}\!{\mathbf{U}}_{{k^d}}^H\!{{\mathbf{H}}_{jk}}\!{{\mathbf{V}}_{{j^u}}}\!{\mathbf{V}}_{{j^u}}^H\!{\mathbf{H}}_{jk}^H\!{{\mathbf{U}}_{{k^d}}}} \!\right)}  \hfill \\
  \quad  - \sum\limits_{k = 1}^K {{\gamma _{{k^u}}}} \sum\limits_{j = 1}^K {{\text{Tr}}\left( {{{\mathbf{W}}_{{k^u}}}{\mathbf{U}}_{{k^u}}^H{{\mathbf{H}}_t}{{\mathbf{V}}_{{j^d}}}{\mathbf{V}}_{{j^d}}^H{\mathbf{H}}_t^H{{\mathbf{U}}_{{k^u}}}} \right)}  \hfill \\
  \quad  - \sum\limits_{k = 1}^K {{\gamma _{{k^u}}}} \sum\limits_{j = 1}^K {{\text{Tr}}\left( {{{\mathbf{W}}_{{k^u}}}{\mathbf{U}}_{{k^u}}^H{{\mathbf{H}}_{{j^u}}}{{\mathbf{V}}_{{j^u}}}{\mathbf{V}}_{{j^u}}^H{\mathbf{H}}_{{j^u}}^H{{\mathbf{U}}_{{k^u}}}} \right)}  + {R_c}, \hfill \\ 
\end{gathered} 
\end{equation}and
\begin{equation}\small
\begin{gathered}
  {R_c} \!=\! \sum\limits_{k = 1}^K {{\gamma _{{k^d}}}} \!\left( {\log \left| {{{\mathbf{W}}_{{k^d}}}} \right| \!- \!{\text{Tr}}\!\left( {{{\mathbf{W}}_{{k^d}}}} \right) \!- \!\!{\text{Tr}}\left( {{{\mathbf{W}}_{{k^d}}}\!{\mathbf{U}}_{{k^d}}^H\!\sigma _{{u_k}}^2{{\mathbf{U}}_{{k^d}}}} \right) \!\!+\! s_d} \right) \hfill \\
 \qquad +\! \sum\limits_{k = 1}^K {{\gamma _{{k^u}}}} \left( \!{\log \left| {{{\mathbf{W}}_{{k^u}}}} \!\right| \!-\! {\text{Tr}}\left(\! {{{\mathbf{W}}_{{k^u}}}} \!\right) \!-\! {\text{Tr}}\left(\! {{{\mathbf{W}}_{{k^u}}}\!{\mathbf{U}}_{{k^u}}^H\!\sigma _r^2\!{{\mathbf{U}}_{{k^u}}}} \right) \!+\! s_u} \right) . \hfill \\
\end{gathered}
\end{equation}
To tackle this problem, we then employ the WMMSE method. Firstly, given ${{\mathbf{V}}_{k^d}}$, ${{\mathbf{V}}_{k^u}}$, ${\mathbf{\Theta}_t}$, ${\mathbf{\Phi}_t}$, ${\mathbf{\Theta}_u}$, ${\mathbf{\Phi}_u}$, we update ${{\mathbf{U}}_{k^d}}$, ${{\mathbf{W}}_{k^d}}$, ${{\mathbf{U}}_{k^u}}$, ${{\mathbf{W}}_{k^u}}$ by using \eqref{U_kd}, \eqref{W_kd}, \eqref{U_ku} and \eqref{W_ku}, respectively. Secondly, for  given ${{\mathbf{U}}_{k^d}}$, ${{\mathbf{W}}_{k^d}}$, ${{\mathbf{U}}_{k^u}}$, ${{\mathbf{W}}_{k^u}}$, ${\mathbf{\Theta}_t}$, ${\mathbf{\Phi}_t}$, ${\mathbf{\Theta}_u}$ and ${\mathbf{\Phi}_u}$, we update ${{\mathbf{V}}_{k^d}}$ and ${{\mathbf{V}}_{k^u}}$. Finally, we update ${\mathbf{\Theta}_t}$, ${\mathbf{\Phi}_t}$, ${\mathbf{\Theta}_u}$ and ${\mathbf{\Phi}_u}$, given ${{\mathbf{U}}_{k^d}}$, ${{\mathbf{W}}_{k^d}}$, ${{\mathbf{U}}_{k^u}}$, ${{\mathbf{W}}_{k^u}}$, ${{\mathbf{V}}_{k^d}}$ and ${{\mathbf{V}}_{k^u}}$.

\subsection{Optimizing the beamforming matrices at the transmitter and users}
For given  amplitudes and phase shifts of the dual-side IOS ${\mathbf{\Theta}_t}$, ${\mathbf{\Phi}_t}$, ${\mathbf{\Theta}_u}$ and ${\mathbf{\Phi}_u}$, the weighted sum rate function can be reformulated as:
\begin{equation}\small
\begin{gathered}
  f\left( {\left. {{{\mathbf{V}}_{{k^d}}},{{\mathbf{V}}_{{k^u}}}} \right|\forall k} \right) \hfill \\
   =  - \sum\limits_{k = 1}^K {{f_1}\left( {{{\mathbf{V}}_{{k^d}}}} \right) \!-\! } \sum\limits_{k = 1}^K {{f_2}\left( {{{\mathbf{V}}_{{k^u}}}} \right)} \! +\! \sum\limits_{k = 1}^K {{\text{Tr}}\left( {{\gamma _{{k^d}}}{{\mathbf{W}}_{{k^d}}}{\mathbf{U}}_{{k^d}}^H{{\mathbf{H}}_{{k^d}}}{{\mathbf{V}}_{{k^d}}}} \right)}  \hfill \\
   \quad + \sum\limits_{k = 1}^K {{\text{Tr}}\left(\! {{\gamma _{{k^d}}}{{\mathbf{W}}_{{k^d}}}\!{\mathbf{V}}_{{k^d}}^H\!{\mathbf{H}}_{{k^d}}^H{{\mathbf{U}}_{{k^d}}}} \!\right)}  \!+\! \sum\limits_{k = 1}^K {{\text{Tr}}\left(\! {{\gamma _{{k^u}}}\!{{\mathbf{W}}_{{k^u}}}{\mathbf{U}}_{{k^u}}^H\!{{\mathbf{H}}_{{k^u}}}{{\mathbf{V}}_{{k^u}}}} \!\right)}  \hfill \\
   \quad + \sum\limits_{k = 1}^K {{\text{Tr}}\left( {{\gamma _{{k^u}}}{{\mathbf{W}}_{{k^u}}}{\mathbf{V}}_{{k^u}}^H{\mathbf{H}}_{{k^u}}^H{{\mathbf{U}}_{{k^u}}}} \right)}  + {R_c}, \hfill \\ 
\end{gathered} 
\end{equation}
where
\begin{equation}\small
\begin{gathered}
  {f_1}\left( {{{\mathbf{V}}_{{k^d}}}} \right) = {\text{Tr}}\left( {{\mathbf{V}}_{{k^d}}^H\left( {{\gamma _{{k^d}}}{\mathbf{H}}_{{k^d}}^H{{\mathbf{U}}_{{k^d}}}{{\mathbf{W}}_{{k^d}}}{\mathbf{U}}_{{k^d}}^H{{\mathbf{H}}_{{k^d}}}} \right.} \right. \hfill \\
 \qquad\qquad\quad  \left. { + \left. {{\mathbf{H}}_t^H\left( {\sum\limits_{j = 1}^K {{\gamma _{{j^u}}}{{\mathbf{U}}_{{j^u}}}{{\mathbf{W}}_{{j^u}}}{\mathbf{U}}_{{j^u}}^H} } \right){{\mathbf{H}}_t}} \right){{\mathbf{V}}_{{k^d}}}} \right), \hfill \\ 
\end{gathered} 
\end{equation}
\begin{equation}\small
\begin{gathered}
  {f_2}\left( {{{\mathbf{V}}_{{k^u}}}} \right) \!=\! {\text{Tr}}\left( \!{{\mathbf{V}}_{{k^u}}^H\!\left( \!{\sum\limits_{j \!= \!1}^K\! {{\gamma _{{j^d}}}\!{\mathbf{H}}_{kj}^H\!{{\mathbf{U}}_{{j^d}}}\!{{\mathbf{W}}_{{j^d}}}\!{\mathbf{U}}_{{j^d}}^H\!{{\mathbf{H}}_{kj}}} }\! \right.} \right. \hfill \\
  \qquad\qquad\quad \left. { +\! \left. {{\mathbf{H}}_{{k^u}}^H\!\left( {\sum\limits_{j\! =\! 1}^K {{\gamma _{{j^u}}}\!{{\mathbf{U}}_{{j^u}}}\!{{\mathbf{W}}_{{j^u}}}\!{\mathbf{U}}_{{j^u}}^H} \!} \right){{\mathbf{H}}_{{k^u}}}}\! \right){{\mathbf{V}}_{{k^u}}}} \right). \hfill \\ 
\end{gathered}   \end{equation}\\
\emph{Proof}: See Appendix A.

By establishing the Lagrangian dual function we have the following expression:
\begin{equation}\small
\begin{gathered}
  \mathcal{L}\left( {\left. {{{\mathbf{V}}_{{k^d}}},{{\mathbf{V}}_{{k^u}}}} \right|\forall k} \right) \hfill \\
   = \sum\limits_{k = 1}^K {{f_1}\left( {{{\mathbf{V}}_{{k^d}}}} \right) \!+\! } \sum\limits_{k = 1}^K {{f_2}\left( {{{\mathbf{V}}_{{k^u}}}} \right)}  - {\text{Tr}}\left( {{\gamma _{{k^d}}}{{\mathbf{W}}_{{k^d}}}{\mathbf{U}}_{{k^d}}^H{{\mathbf{H}}_{{k^d}}}{{\mathbf{V}}_{{k^d}}}} \right) \hfill \\
  \quad  -\! {\text{Tr}}\left( {{\gamma _{{k^d}}}{{\mathbf{W}}_{{k^d}}}{\mathbf{V}}_{{k^d}}^H{\mathbf{H}}_{{k^d}}^H{{\mathbf{U}}_{{k^d}}}} \right) \!-\! {\text{Tr}}\left( {{\gamma _{{k^u}}}{{\mathbf{W}}_{{k^u}}}\!{\mathbf{U}}_{{k^u}}^H\!{{\mathbf{H}}_{{k^u}}}{{\mathbf{V}}_{{k^u}}}} \right) \hfill \\
 \quad   - {\text{Tr}}\left( {{\gamma _{{k^u}}}\!{{\mathbf{W}}_{{k^u}}}{\mathbf{V}}_{{k^u}}^H{\mathbf{H}}_{{k^u}}^H{{\mathbf{U}}_{{k^u}}}} \right) \!+\! \sum\limits_{k = 1}^K {{\lambda _{{k^u}}}\left( {{\text{Tr}}\left(\! {{{\mathbf{V}}_{{k^u}}}{\mathbf{V}}_{{k^u}}^H} \right)\! - \!{P_U}} \!\right)}  \hfill \\
 \quad   + {\mu _d}\left( {\sum\limits_{k = 1}^K \!{{\text{Tr}}\!\left( {{{\mathbf{V}}_{{k^d}}}\!{\mathbf{V}}_{{k^d}}^H} \right) - {P_B}} } \right), \hfill \\ 
\end{gathered} 
\end{equation}
where $\lambda _{{k^u}}$ and $\mu _d$ are the Lagrangian dual factors for the $k$th uplink and downlink transmissions, respectively.

Therefore, by exploiting the first-order derivative of the functions regarding  the transmit beamforming matrices at the transmitter and users $\mathbf{V}_{k^d}$ and $\mathbf{V}_{k^u}$, then we have
\begin{equation}\small
\begin{gathered}
  \frac{{\partial \mathcal{L}\left( {{{\mathbf{V}}_{{k^d}}}} \right)}}{{\partial {{\mathbf{V}}_{{k^d}}}}} = 2\left( \begin{gathered}
  {\gamma _{{k^d}}}{\mathbf{H}}_{{k^d}}^H{{\mathbf{U}}_{{k^d}}}{{\mathbf{W}}_{{k^d}}}{\mathbf{U}}_{{k^d}}^H{{\mathbf{H}}_{{k^d}}} + {\mu _d}{\mathbf{I}} \hfill \\
   + {\mathbf{H}}_t^H\left( {\sum\limits_{j = 1}^K {{\gamma _{{j^u}}}} {{\mathbf{U}}_{{j^u}}}{{\mathbf{W}}_{{j^u}}}{\mathbf{U}}_{{j^u}}^H} \right){{\mathbf{H}}_t} \hfill \\ 
\end{gathered}  \right){{\mathbf{V}}_{{k^d}}} \hfill \\
 \qquad  \qquad\qquad\quad - 2{\gamma _{{k^d}}}{\mathbf{H}}_{{k^d}}^H{{\mathbf{U}}_{{k^d}}}{{\mathbf{W}}_{{k^d}}}, \hfill \\ 
\end{gathered} 
\end{equation}
\begin{equation}\small
\begin{gathered}
  \frac{{\partial \mathcal{L}\left( {{{\mathbf{V}}_{{k^u}}}} \right)}}{{\partial {{\mathbf{V}}_{{k^u}}}}} = 2\left( \begin{gathered}
  \left( {\sum\limits_{j = 1}^K {{\gamma _{{j^d}}}{\mathbf{H}}_{kj}^H{{\mathbf{U}}_{{j^d}}}{{\mathbf{W}}_{{j^d}}}{\mathbf{U}}_{{j^d}}^H{{\mathbf{H}}_{kj}}}  + {\lambda _{{k^u}}}{\mathbf{I}}} \right. \hfill \\
   + {\mathbf{H}}_{{k^u}}^H\left( {\sum\limits_{j = 1}^K {{\gamma _{{j^u}}}} {{\mathbf{U}}_{{j^u}}}{{\mathbf{W}}_{{j^u}}}{\mathbf{U}}_{{j^u}}^H} \right){{\mathbf{H}}_{{k^u}}} \hfill \\ 
\end{gathered}  \right){{\mathbf{V}}_{{k^u}}} \hfill \\
 \qquad  \qquad \qquad\quad    - 2{\gamma _{{k^u}}}{\mathbf{H}}_{{k^u}}^H{{\mathbf{U}}_{{k^u}}}{{\mathbf{W}}_{{k^u}}}, \hfill \\ 
\end{gathered}  
\end{equation}
\begin{algorithm}[!t]
\caption{Bisection Search}
\label{randomization_bis}
\algsetup{linenosize=\footnotesize}
\footnotesize
\begin{algorithmic}
\STATE 1. Initialize $0 \leq{\lambda _{k^u}^l} \leq{\lambda_{k^u} ^u}$, $0 \leq{\mu _{d}^l} \leq{\lambda_{d} ^u}$, $\forall k$ and the accuracy $\varepsilon_b$. \\
\STATE 2. $\textbf{while}$ $\left| {{\lambda _{k^u}^u} - {\lambda _{k^u}^l}} \right| > \varepsilon_b $ and $\left| {{\mu _{d}^u} - {\mu _{d}^l}} \right| > \varepsilon_b $ $\textbf{do}$: \\
\STATE 3. \quad Compute $\lambda_k  = \frac{{{\lambda _{k^u} ^l} + {\lambda _{k^u} ^u}}}{2}$,  $\mu_d  = \frac{{{\mu _{d} ^l} + {\mu_{d} ^u}}}{2}$,  ${\mathbf{V}}_{k^d}$, ${\mathbf{V}}_{k^u}$ of \eqref{V_kd} and \eqref{V_ku}, respectively, then  evaluate ${\text{Tr}}\left( {{{\mathbf{V}}_{{k^d}}}{\mathbf{V}}_{{k^d}}^H} \right)$ and ${\text{Tr}}\left( {{{\mathbf{V}}_{{k^u}}}{\mathbf{V}}_{{k^u}}^H} \right)$.
\STATE 4.     \quad \textbf{if} $\sum\limits_{k = 1}^K {{\text{Tr}}\left( {{{\mathbf{V}}_{{k^d}}}{\mathbf{V}}_{{k^d}}^H} \right)}  \geq {P_B}$ and ${\text{Tr}}\left( {{{\mathbf{V}}_{{k^u}}}{\mathbf{V}}_{{k^u}}^H} \right) \geq {P_U}$
\STATE 5. \qquad ${\lambda_{k^u} ^l} = \lambda_{k^u}$ and ${\mu_{d} ^l} = \mu_{d}$.
\STATE 6.  \quad \textbf{else}
\STATE 7.  \qquad ${\lambda_{k^u} ^u} = \lambda_{k^u}$ and ${\mu_{d} ^u} = \mu_{d}$.
\STATE 8.  \quad \textbf{end}
\STATE 9.   $\textbf{end}$ $\textbf{while}$
\end{algorithmic}
\end{algorithm}by setting them to zeros, we have the optimal solutions as follows:
\begin{equation}\small\label{V_kd}
{\mathbf{V}}_{{k^d}}^* = {\mathbf{\Xi}}_{k^d}^{ - 1}{\gamma _{k^d}}{\mathbf{H}}_{k^d}^H{{\mathbf{U}}_{k^d}}{{\mathbf{W}}_{k^d}},
\end{equation}
\begin{equation}\small\label{V_ku}
{\mathbf{V}}_{{k^u}}^* = {\mathbf{\Xi}}_{k^u}^{ - 1}{\gamma _{k^u}}{\mathbf{H}}_{k^u}^H{{\mathbf{U}}_{k^u}}{{\mathbf{W}}_{k^u}},
\end{equation}
where
\begin{equation}\small
\begin{gathered}
  {{\mathbf{\Xi }}_{{k^d}}} = {\gamma _{{k^d}}}{\mathbf{H}}_{{k^d}}^H{{\mathbf{U}}_{{k^d}}}{{\mathbf{W}}_{{k^d}}}{\mathbf{U}}_{{k^d}}^H{{\mathbf{H}}_{{k^d}}} + {\mu _d}{\mathbf{I}} \hfill \\
 \qquad\quad   + {\gamma _{{k^u}}}{\mathbf{H}}_t^H\left( {\sum\limits_{j = 1}^K {{{\mathbf{U}}_{{j^u}}}{{\mathbf{W}}_{{j^u}}}{\mathbf{U}}_{{j^u}}^H} } \right){{\mathbf{H}}_t}, \hfill \\ 
\end{gathered} 
\end{equation}
\begin{equation}\small
\begin{gathered}
  {{\mathbf{\Xi }}_{{k^u}}} = \sum\limits_{j = 1}^K {{\gamma _{{j^d}}}{\mathbf{H}}_{kj}^H{{\mathbf{U}}_{{j^d}}}{{\mathbf{W}}_{{j^d}}}{\mathbf{U}}_{{j^d}}^H{{\mathbf{H}}_{kj}}}  + {\lambda _{{k^u}}}{\mathbf{I}} \hfill \\
 \qquad \quad + {\mathbf{H}}_{{k^u}}^H\left( {\sum\limits_{j = 1}^K {{\gamma _{{j^u}}}{{\mathbf{U}}_{{j^u}}}{{\mathbf{W}}_{{j^u}}}{\mathbf{U}}_{{j^u}}^H} } \right){{\mathbf{H}}_{{k^u}}}. \hfill \\ 
\end{gathered} 
\end{equation}
Note that ${{\lambda _{k^u}}}$ and ${{\mu _d}}$ are independent of each other and each can be found using bisection search to guarantee ${\text{Tr}}\left( {{{\mathbf{V}}_{{k^u}}}{\mathbf{V}}_{{k^u}}^H} \right) \leq {P_U}$ and $\sum\limits_{k = 1}^K {{\text{Tr}}\left( {{{\mathbf{V}}_{{k^d}}}{\mathbf{V}}_{{k^d}}^H} \right) \leq {P_B}}$, respectively.  It is important to identify whether $\mu_d=0$ and $\lambda_{k^u}=0$ satisfy the total power constraint at the transmitter, and the $k$th user, respectively, before starting the bisection search. In addition,  the details of the bisection search method are provided in Algorithm \ref{randomization_bis}.

\subsection{Optimizing the reflecting and refracting phase shift matrices of the dual-side IOS}
Given the beamforming matrices of downlink and uplink transmissions $\mathbf{V}_{k^d}$ and $\mathbf{V}_{k^u}$, the weighted sum rate function can be reformulated as:
\begin{equation}\small
\label{g_theta_phi}
\begin{gathered}
  g\left( {{{\mathbf{\Theta }}_t},{{\mathbf{\Phi }}_t},{{\mathbf{\Theta }}_u},{{\mathbf{\Phi }}_u}} \right) \hfill \\
   =  - \sum\limits_{k = 1}^K {{\text{Tr}}\left( {{\mathbf{\Phi }}_t^H{{\mathbf{A}}_k}{{\mathbf{\Phi }}_t}{{\mathbf{B}}_k}} \right)}  + \sum\limits_{k = 1}^K {{\text{Tr}}\left( {{\mathbf{\Phi }}_t^H{\mathbf{C}}_k^H} \right)}  + \sum\limits_{k = 1}^K {{\text{Tr}}\left( {{{\mathbf{\Phi }}_t}{{\mathbf{C}}_k}} \right)}  \hfill \\
  \quad   - \sum\limits_{k = 1}^K {\sum\limits_{j = 1}^K {{\text{Tr}}\left( {{\mathbf{\Theta }}_t^H{{\mathbf{A}}_k}{{\mathbf{\Phi }}_t}{{\mathbf{D}}_j}} \right)} }  + \sum\limits_{k = 1}^K {\sum\limits_{j = 1}^K {{\text{Tr}}\left( {{\mathbf{\Theta }}_t^H{\mathbf{F}}_{kj}^H} \right)} }  \hfill \\
  \quad   + \sum\limits_{k = 1}^K {\sum\limits_{j = 1}^K {{\text{Tr}}\left( {{{\mathbf{\Theta }}_t}{{\mathbf{F}}_{kj}}} \right)} }  - \sum\limits_{k = 1}^K {\sum\limits_{j = 1}^K {{\text{Tr}}\left( {{\mathbf{\Theta }}_u^H{{\mathbf{X}}_k}{{\mathbf{\Theta }}_u}{{\mathbf{B}}_j}} \right)} }  \hfill \\
  \quad   + \sum\limits_{k = 1}^K {\sum\limits_{j = 1}^K {{\text{Tr}}\left( {{\mathbf{\Theta }}_u^H{\mathbf{Y}}_{kj}^H} \right)} }  + \sum\limits_{k = 1}^K {\sum\limits_{j = 1}^K {{\text{Tr}}\left( {{{\mathbf{\Theta }}_u}{{\mathbf{Y}}_{kj}}} \right)} }  \hfill \\
  \quad   + \sum\limits_{k = 1}^K {{\text{Tr}}\left( {{\mathbf{\Phi }}_u^H{\mathbf{Z}}_k^H} \right) + \sum\limits_{k = 1}^K {{\text{Tr}}\left( {{{\mathbf{\Phi }}_u}{{\mathbf{Z}}_k}} \right)} }  \hfill \\
 \quad    - \sum\limits_{k = 1}^K {\sum\limits_{j = 1}^K {{\text{Tr}}\left( {{\mathbf{\Phi }}_u^H{{\mathbf{X}}_k}{{\mathbf{\Phi }}_u}{{\mathbf{D}}_j}} \right)} }  + {R_{cg}}. \hfill \\ 
\end{gathered}  
\end{equation}
\emph{The derivation of \eqref{g_theta_phi} and ${\mathbf{A}}_k$, ${\mathbf{B}}_k$, ${\mathbf{C}}_k$, ${\mathbf{D}}_j$, ${\mathbf{F}}_{kj}$, ${\mathbf{X}}_k$, ${\mathbf{Y}}_{kj}$, ${\mathbf{Z}}_k$ and $R_{cg}$}: Can be found in Appendix B.

The weighted sum rate maximization problem with respect to the reflecting and refracting phase shifts is equivalent to:

\vspace{-0.4em}
\begin{small}
\begin{align}
\mathop {\max }\limits_{\mathbf{\Theta}_t,\mathbf{\Phi}_t,\mathbf{\Theta}_u,\mathbf{\Phi}_u}&\ \ g\left( {{{\mathbf{\Theta }}_t},{{\mathbf{\Phi }}_t},{{\mathbf{\Theta }}_u},{{\mathbf{\Phi }}_u}} \right)  \label{P3_OF}\\
{\text{s.t.}}\ \ & \ \   a_{dir,l}^2 + b_{dir,l}^2 \leq 1,  \tag{\ref{P3_OF}{a}}  \label{P3_modulus1} \\
& \ \ 0 \leq{a_{dir,l}},{b_{dir,l}} \leq 1, \tag{\ref{P3_OF}{b}}  \label{P3_modulus2}   \\
& \ \  0 \leq {\alpha _{dir,l}},{\beta _{dir,l}} \leq 2\pi , \ \  \forall l. \tag{\ref{P3_OF}{c}}  \label{P3_modulus3}
\end{align}\end{small}
However, the problem \eqref{P3_OF} is still non-convex.  By utilizing the matrix properties in \cite{Matrix_Analysis}, and by defining ${{\boldsymbol{\theta }}_t} = {\left[ {{a_{t,1}}{e^{j{\alpha _{t,1}}}},{a_{t,2}}{e^{j{\alpha _{t,2}}}}, \ldots ,{a_{t,L}}{e^{j{\alpha _{t,L}}}}} \right]^T} \in {\mathbb{C}^{L \times 1}}$, ${\boldsymbol{\phi} _t} = {\left[ {{b_{t,1}}{e^{j{\beta _{t,1}}}},{b_{t,2}}{e^{j{\beta _{t,2}}}}, \ldots ,{b_{t,L}}{e^{j{\beta _{t,L}}}}} \right]^T} \in {\mathbb{C}^{L \times 1}}$, ${{\boldsymbol{\theta }}_u}\! \!=\! {\left[ {{a_{u,1}}{e^{j{\alpha _{u,1}}}},{a_{u,2}}{e^{j{\alpha _{u,2}}}}, \ldots ,{a_{u,L}}{e^{j{\alpha _{u,L}}}}} \right]^T} $ $\in {\mathbb{C}^{L \times 1}}$ and ${\boldsymbol{\phi} _u} \!=\!\! {\left[ {{b_{u,1}}{e^{j{\beta _{u,1}}}},{b_{u,2}}{e^{j{\beta _{u,2}}}}, \ldots ,{b_{u,L}}{e^{j{\beta _{u,L}}}}} \right]^T} \in {\mathbb{C}^{L \times 1}}$, we have

\begin{small}
\begin{equation*}
\sum\limits_{k = 1}^K {{\text{Tr}}\left( {{\mathbf{\Phi }}_t^H{{\mathbf{A}}_k}{{\mathbf{\Phi }}_t}{{\mathbf{B}}_k}} \right)} \! =\! \phi _t^H\left( {\sum\limits_{k = 1}^K {{{\mathbf{A}}_k}\! \odot \! {{\mathbf{B}}_k}} } \right){\phi _t},  
\end{equation*}\end{small}
\begin{small}
\begin{equation*}
 \sum\limits_{k = 1}^K {\sum\limits_{j = 1}^K {{\text{Tr}}\left( {{\mathbf{\Theta }}_t^H\!{{\mathbf{A}}_k}\!{{\mathbf{\Phi }}_t}\!{{\mathbf{D}}_j}} \right)} }  = {\mathbf{\theta }}_t^H\! \left( {\sum\limits_{k = 1}^K {\sum\limits_{j = 1}^K {{{\mathbf{A}}_k} \! \odot {{\mathbf{D}}_j}} } } \right)\!{{\mathbf{\theta }}_t},
\end{equation*}\end{small}
\begin{small}
\begin{equation*}
\sum\limits_{k = 1}^K \!{\sum\limits_{j = 1}^K {{\text{Tr}}\left( {{\mathbf{\Phi }}_u^H\!{{\mathbf{X}}_k}\!{{\mathbf{\Phi }}_u}\!{{\mathbf{D}}_j}} \right)} }  \!=\! \phi _u^H\left( {\sum\limits_{k = 1}^K\! {\sum\limits_{j = 1}^K \!{{{\mathbf{X}}_k} \!\odot\! {{\mathbf{D}}_j}} } } \right)\!{\phi _u},  
\end{equation*}\end{small}
\begin{small}
\begin{equation*}
 \sum\limits_{k = 1}^K {\sum\limits_{j = 1}^K {{\text{Tr}}\left( {{\mathbf{\Theta }}_u^H{{\mathbf{X}}_k}\! {{\mathbf{\Theta }}_u}\! {{\mathbf{B}}_j}} \right)} } \!  = \! {\mathbf{\theta }}_u^H\! \left( {\sum\limits_{k = 1}^K {\sum\limits_{j = 1}^K {{{\mathbf{X}}_k} \! \odot\!  {{\mathbf{B}}_j}} } } \right)\! {{\mathbf{\theta }}_u},
\end{equation*}\end{small}
\begin{small}
\begin{equation*}
\sum\limits_{k = 1}^K {{\text{Tr}}\left( {{\mathbf{\Phi }}_t^H{\mathbf{C}}_k^H} \right)}  \! +\!  \sum\limits_{k = 1}^K {{\text{Tr}}\left( {{{\mathbf{\Phi }}_t}{{\mathbf{C}}_k}} \right)}  \! = \! 2\operatorname{Re} \left\{ {\phi _t^H{{\mathbf{c}}^*}} \right\}, 
\end{equation*}\end{small}
\begin{small}
\begin{equation*}
 \sum\limits_{k = 1}^K {\sum\limits_{j = 1}^K {{\text{Tr}}\left( {{\mathbf{\Theta }}_t^H{\mathbf{F}}_{kj}^H} \right)} } \!  +\!  \sum\limits_{k = 1}^K {\sum\limits_{j = 1}^K {{\text{Tr}}\left( {{{\mathbf{\Theta }}_t}{{\mathbf{F}}_{kj}}} \right)} } \!  = \! 2\operatorname{Re} \left\{ {{\mathbf{\theta }}_t^H{{\mathbf{f}}^*}} \right\},
\end{equation*}\end{small}
\begin{small}
\begin{equation*}
\sum\limits_{k = 1}^K {{\text{Tr}}\left( {{\mathbf{\Phi }}_u^H{\mathbf{Z}}_k^H} \right)\!  +\!  \sum\limits_{k = 1}^K {{\text{Tr}}\left( {{{\mathbf{\Phi }}_u}{{\mathbf{Z}}_k}} \right)} } \!  = \! 2\operatorname{Re} \left\{ {\phi _u^H{{\mathbf{z}}^*}} \right\}, 
\end{equation*}\end{small}
\begin{small}
\begin{equation*}
 \sum\limits_{k = 1}^K {\sum\limits_{j = 1}^K {{\text{Tr}}\left( {{\mathbf{\Theta }}_u^H{\mathbf{Y}}_{kj}^H} \right)} }  \! +\!  \sum\limits_{k = 1}^K {\sum\limits_{j = 1}^K {{\text{Tr}}\left( {{{\mathbf{\Theta }}_u}{{\mathbf{Y}}_{kj}}} \right)} } \!  =\!  2\operatorname{Re} \left\{ {{\mathbf{\theta }}_u^H{{\mathbf{y}}^*}} \right\}.
\end{equation*}\end{small}
\begin{algorithm}[!t]
\caption{Weighted sum rate maximization for the proposed dual-side IOS assisted system}\label{wsr_algo}
\label{Maximizing}
\algsetup{linenosize=\footnotesize}
\footnotesize
\begin{algorithmic}
\STATE {Initialize the beamforming matrices of downlink and uplink transmissions ${\mathbf{V}}_{k^d}^0$, ${\mathbf{V}}_{k^u}^0$, $\forall k$, and the reflecting, refracting amplitudes and phase shifts ${\mathbf{\Theta}_t ^{0}}$, ${\mathbf{\Phi}_t ^{0}}$, ${\mathbf{\Theta}_u ^{0}}$, ${\mathbf{\Phi}_u ^{0}}$.
Compute $R_{u^k}^d\left( {{\mathbf{V}}_{k^d}^0,{\mathbf{V}}_{k^u}^0,\!{{\mathbf{\Theta }}_t^0},\!{{\mathbf{\Phi }}_t^0}}, {{\mathbf{\Theta }}_u^0} ,{{\mathbf{\Phi }}_u^0}  \right)$ and $R_{u^k}^u\left( {{\mathbf{V}}_{k^d}^0,{\mathbf{V}}_{k^u}^0,\!{{\mathbf{\Theta }}_t^0},\!{{\mathbf{\Phi }}_t^0}}, {{\mathbf{\Theta }}_u^0} ,{{\mathbf{\Phi }}_u^0}  \right)$ using \eqref{R_ud} and \eqref{R_uu}, respectively. Then evaluate the objective function in \eqref{P1_OF} as  $R_{w}^0\left( {{\mathbf{V}}_{k^d}^0,{\mathbf{V}}_{k^u}^0,\!{{\mathbf{\Theta }}_t^0},\!{{\mathbf{\Phi }}_t^0}}, {{\mathbf{\Theta }}_u^0} ,{{\mathbf{\Phi }}_u^0}  \right)$ , set $t=0$ and the accuracy for iteration $\varepsilon_{w}$.}
\REPEAT
\STATE  1. Given  ${\mathbf{V}}_{k^d}^t$, ${\mathbf{V}}_{k^u}^t$, ${\mathbf{\Theta}_t ^t}$, ${\mathbf{\Phi}_t ^{t}}$, ${\mathbf{\Theta}_u ^{t}}$, ${\mathbf{\Phi}_u ^{t}}$.
Compute the $k$th rate of the downlink and uplink transmission as  $R_{u^k}^d\left( {{\mathbf{V}}_{k^d}^t,{\mathbf{V}}_{k^u}^t,\!{{\mathbf{\Theta }}_t^t},\!{{\mathbf{\Phi }}_t^t}}, {{\mathbf{\Theta }}_u^t}, {{\mathbf{\Phi }}_u^t}  \right)$ and $R_{u^k}^u\left( {{\mathbf{V}}_{k^d}^t,{\mathbf{V}}_{k^u}^t,\!{{\mathbf{\Theta }}_t^t},\!{{\mathbf{\Phi }}_t^t}}, {{\mathbf{\Theta }}_u^t}, {{\mathbf{\Phi }}_u^t}  \right)$, respectively.\\
2. Evaluate $\mathbf{U}_{k^d}^t$, $\mathbf{W}_{k^d}^t$, $\mathbf{U}_{k^u}^t$ and $\mathbf{W}_{k^u}^t$ by using \eqref{U_kd}, \eqref{W_kd}, \eqref{U_ku} and \eqref{W_ku}, respectively.\\
3. Given $\mathbf{U}_{k^d}^t$, $\mathbf{W}_{k^d}^t$, $\mathbf{U}_{k^u}^t$, $\mathbf{W}_{k^u}^t$, the reflecting and refracting phase shifts ${\mathbf{\Theta}_t ^{t}}$, ${\mathbf{\Phi}_t ^{t}}$, ${\mathbf{\Theta}_u ^{t}}$ and ${\mathbf{\Phi}_u ^{t}}$, calculate the beamforming matrices of downlink and uplink transmissions ${\mathbf{V}}_{k^d}^{t+1}$, ${\mathbf{V}}_{k^u}^{t+1}$ by utilizing \eqref{V_kd} and \eqref{V_ku}, and obtain the optimal Lagrangian dual factors via the bisection search method.  \\
4. Given $\mathbf{U}_{k^d}^t$, $\mathbf{W}_{k^d}^t$, $\mathbf{U}_{k^u}^t$, $\mathbf{W}_{k^u}^t$, the beamforming matrices ${\mathbf{V}}_{k^d}^{t+1}$ and ${\mathbf{V}}_{k^u}^{t+1}$, calculate the reflecting and refracting phase shifts  ${\mathbf{\Theta}_t ^{t+1}}$, ${\mathbf{\Phi}_t ^{t+1}}$, ${\mathbf{\Theta}_u ^{t+1}}$ and ${\mathbf{\Phi}_u ^{t+1}}$   by utilizing the QCQP algorithm.  \\
5. Compute the rates of downlink and uplink transmissions as $R_{u^k}^d\left( {{\mathbf{V}}_{k^d}^{t+1},{\mathbf{V}}_{k^u}^{t+1},\!{{\mathbf{\Theta }}_t^{t+1}},\!{{\mathbf{\Phi }}_t^{t+1}}}, {{\mathbf{\Theta }}_u^{t+1}} ,{{\mathbf{\Phi }}_u^{t+1}}  \right)$ and $R_{u^k}^u\left( {{\mathbf{V}}_{k^d}^{t+1},{\mathbf{V}}_{k^u}^{t+1},\!{{\mathbf{\Theta }}_t^{t+1}},\!{{\mathbf{\Phi }}_t^{t+1}}}, {{\mathbf{\Theta }}_u^{t+1}} ,{{\mathbf{\Phi }}_u^{t+1}}  \right)$ using \eqref{R_ud} and \eqref{R_uu}, then calculate the weighted sum rate using as \eqref{P1_OF} $R_w^{t+1}$.\\
6. Set $t = t+1$.
\UNTIL {$\frac{{\left| R_{w}^{t+1}\left(\!\! {{\mathbf{V}}_{k^d}^{t+1},\!\!{\mathbf{V}}_{k^u}^{t+1},\!\!{{\mathbf{\Theta }}_t^{t+1}},\!{{\mathbf{\Phi }}_t^{t+1}}},\! {{\mathbf{\Theta }}_u^{t+1}} ,\!{{\mathbf{\Phi }}_u^{t+1}} \!\! \right) \!\!-\!\! R_{w}^t\left(\! {{\mathbf{V}}_{k^d}^t,\!{\mathbf{V}}_{k^u}^t,\!{{\mathbf{\Theta }}_t^t},\!{{\mathbf{\Phi }}_t^t}},\! {{\mathbf{\Theta }}_u^t} ,\!{{\mathbf{\Phi }}_u^t}  \!\right) \right|}}{R_{w}^{t+1}\left(\! {{\mathbf{V}}_{k^d}^{t+1},\!{\mathbf{V}}_{k^u}^{t+1},\!{{\mathbf{\Theta }}_t^{t+1}},\!{{\mathbf{\Phi }}_t^{t+1}}}, \!{{\mathbf{\Theta }}_u^{t+1}} ,\!{{\mathbf{\Phi }}_u^{t+1}}\! \! \right)}$ $ \leq \varepsilon_w$ .}
\end{algorithmic}
\end{algorithm}

In addition, the original constraints can be transformed to $\text{diag}\left\{ {{\boldsymbol{\phi} _t}\boldsymbol{\phi} _t^H + {{\boldsymbol{\theta }}_t}{\boldsymbol{\theta }}_t^H} \right\} \leq {{\mathbf{1}}_M}$ and $\text{diag}\left\{ {{\boldsymbol{\phi} _u}\boldsymbol{\phi} _u^H + {{\boldsymbol{\theta }}_u}{\boldsymbol{\theta }}_u^H} \right\} \leq {{\mathbf{1}}_M}$, then problem \eqref{P3_OF} is equivalent to

\vspace{-0.4em}
\begin{small}
\begin{align}
\mathop {\min }\limits_{\boldsymbol{\phi}_t,\boldsymbol{\theta}_t,\boldsymbol{\phi}_u,\boldsymbol{\theta}_u}&\ \ g'\left( \boldsymbol{\phi}_t,\boldsymbol{\theta}_t,\boldsymbol{\phi}_u,\boldsymbol{\theta}_u \right) \label{P4_OF}\\
{\text{s.t.}}\ \ & \ \   \text{diag}\left\{ {{\boldsymbol{\phi} _t}\boldsymbol{\phi} _t^H + {{\boldsymbol{\theta }}_t}{\boldsymbol{\theta }}_t^H} \right\} \leq {{\mathbf{1}}_M}  \tag{\ref{P4_OF}{a}}  \label{P4_modulus1} \\
& \ \  \text{diag} \left\{ {{\boldsymbol{\phi} _u}\boldsymbol{\phi} _u^H + {{\boldsymbol{\theta }}_u}{\boldsymbol{\theta }}_u^H} \right\} \leq {{\mathbf{1}}_M}, \tag{\ref{P4_OF}{b}}  \label{P4_modulus2}
\end{align}\end{small}where
\begin{equation}\small
\begin{gathered}
  g'\left( {{\phi _t},{{\mathbf{\theta }}_t},{\phi _u},{{\mathbf{\theta }}_u}} \right) \hfill \\
   = \phi _t^H\left( {\sum\limits_{k = 1}^K {{{\mathbf{A}}_k} \odot {{\mathbf{B}}_k}} } \right){\phi _t} + {\mathbf{\theta }}_t^H\left( {\sum\limits_{k = 1}^K {\sum\limits_{j = 1}^K {{{\mathbf{A}}_k} \odot {{\mathbf{D}}_j}} } } \right){{\mathbf{\theta }}_t} \hfill \\
  \quad - 2\operatorname{Re} \left\{ {\phi _t^H{{\mathbf{c}}^*}} \right\} - 2\operatorname{Re} \left\{ {{\mathbf{\theta }}_t^H{{\mathbf{f}}^*}} \right\} + \phi _u^H\left( {\sum\limits_{k = 1}^K {\sum\limits_{j = 1}^K {{{\mathbf{X}}_k} \odot {{\mathbf{D}}_j}} } } \right){\phi _u} \hfill \\
 \quad  + {\mathbf{\theta }}_u^H\left( {\sum\limits_{k = 1}^K {\sum\limits_{j = 1}^K {{{\mathbf{X}}_k} \odot {{\mathbf{B}}_j}} } } \right){{\mathbf{\theta }}_u} - 2\operatorname{Re} \left\{ {\phi _u^H{{\mathbf{z}}^*}} \right\} - 2\operatorname{Re} \left\{ {{\mathbf{\theta }}_u^H{{\mathbf{y}}^*}} \right\}, \hfill \\ 
\end{gathered} 
\end{equation}
and
${\mathbf{c}} = \text{diag} \left\{ {\sum\limits_{k = 1}^K {{{\mathbf{C}}_k}} } \right\}$, ${\mathbf{f}} =\text{diag} \left\{ {\sum\limits_{k = 1}^K  \sum\limits_{j = 1}^K {{{\mathbf{F}}_{kj}}} } \right\}$, ${\mathbf{y}} = \text{diag} \left\{ {\sum\limits_{k = 1}^K \sum\limits_{j = 1}^K {{{\mathbf{Y}}_{kj}}} } \right\}$ and ${\mathbf{z}} =\text{diag}\left\{ {\sum\limits_{k = 1}^K {{{\mathbf{Z}}_k}} } \right\}$. Since the objective function in (39) is a quadratically convex expression for the minimization problem, and the constraints are also quadratically convex, then this problem can be solved using the QCQP method \cite{convexopt}.

\subsection{Overall algorithm and complexity analysis}
Overall, we provide an iterative algorithm that aims at maximizing the weighted sum rate as in Algorithm \ref{wsr_algo}.
The Lagrangian dual approach is leveraged to optimize the beamforming matrices of the transmitter and multiple users. 
Furthermore, the QCQP procedure is employed for optimizing both amplitudes, phase shifts for reflection and refraction of both sides of the  IOS.
Each step in Algorithm \ref{wsr_algo} ensures the objective function increases monotonically, i.e., $R_w^t\left( {{\mathbf{V}}_{{k^d}}^t,{\mathbf{V}}_{{k^u}}^t,{\mathbf{\Theta }}_t^t,{\mathbf{\Phi }}_t^t,{\mathbf{\Theta }}_u^t,{\mathbf{\Phi }}_u^t} \right) >  \ldots  > R_w^0\left( {{\mathbf{V}}_{{k^d}}^0,{\mathbf{V}}_{{k^u}}^0,{\mathbf{\Theta }}_t^0,{\mathbf{\Phi }}_t^0,{\mathbf{\Theta }}_u^0,{\mathbf{\Phi }}_u^0} \right)$,  therefore, the proposed algorithm's convergence is guaranteed.
Moreover, as for the beamforming and phase shifts optimization, the computational complexities of the Lagrangian dual method and QCQP method are $\mathcal{O}\left( {KN_t^3 + KN_{ut}^3} \right)$ and $\mathcal{O}\left( {{L^3}} \right)$ \cite{convexopt}, respectively.
Thus, the total complexity order of the proposed algorithm amounts to $\mathcal{O}\left( {{N_{ite}} \times \max \left\{ {KN_t^3 + KN_t^3,{L^3}} \right\}} \right)$, where $N_{ite}$ is the number of optimization iterations\footnote{In addition, algorithms with low computational complexity for such networks will be developed in one of our future works.}. 

\vspace{1em}
\section{Numerical Results}
This section provides the settings for simulations and demonstrates the performance of the proposed algorithms and methods.
In the simulation, the number of users $K = 3$, and the corresponding 3-dimensional coordinates are $\left[ {20,20,1.5} \right]$, $\left[ {25,- 35,1.5} \right]$ and $\left[ {35, - 25,1.5} \right]$ m, respectively. 
In addition, we set $\left[ {0,0,5} \right]$, $\left[ {0,1,5} \right]$, $\left[ {1,1,5} \right]$ m as the locations of the typical transmit antenna, receive antenna, and element at the dual-side IOS, respectively.
$\frac{\lambda }{2}$ is the interval between any two neighboring elements or antennas, where $\lambda=0.05$ m. Also, 
$\kappa=2.5$ is the Rician exponent and $\chi=3 $ dB is the Rician factor, respectively.
Furthermore, $\sigma_{u_k}=\sigma_r=-80$ dBm are the noise powers of $k$th user and the transmitter, respectively, and $\varepsilon_w  =\varepsilon_b= {10^{ - 4}}$ are convergence measurement values.
The  weighted factors for the $k$th downlink and uplink transmissions are $[0.746,0.560,0.781]$ and $[0.328,0.340,0.595]$, respectively.
Specifically, the performance in the DS-IOS system using the proposed  algorithms is compared with the following benchmark schemes:
\begin{enumerate}
\item \textbf{WO-IOS}: 
For this case, no IOS is introduced and only the beamforming matrices for uplink and downlink signals are optimized, which is selected as the benchmark for performance comparison.  

\item \textbf{SS-IOS}: The single-side IOS, a.k.a. simultaneously transmitting and reflecting (STAR)-RIS, is employed in our system for comparison; specifically, only the uplink transmission from the users to the transmitter is considered. Besides, the beamforming matrices of the uplink transmission and the reflection and refraction phase shifts at the user side are optimized.

\end{enumerate}
\begin{figure}[t]
        \centering
        \includegraphics*[width=80mm]{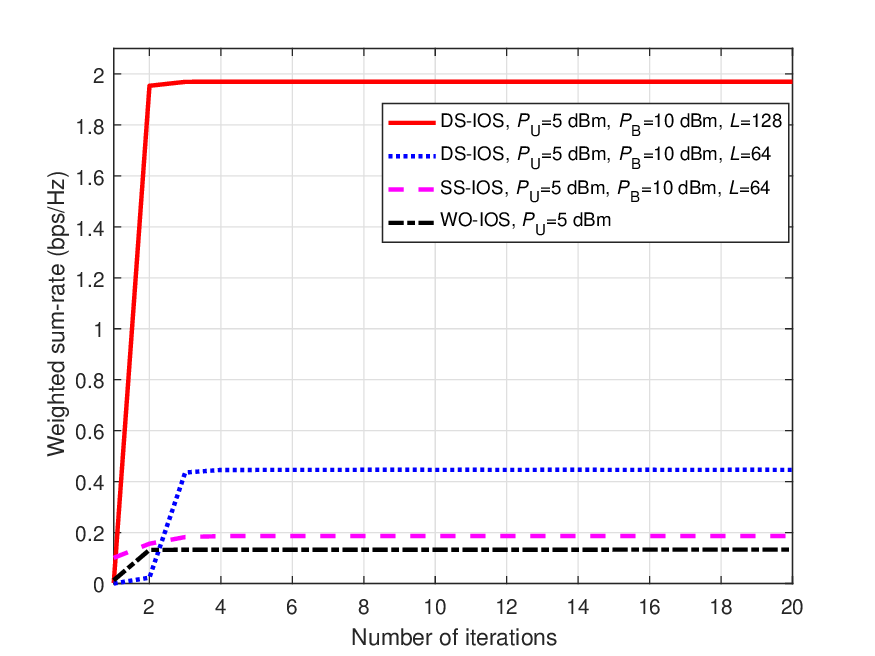}
       \caption{Iteration behaviors of the formulated problem for different schemes.}
        \label{fig2}
        \vspace{-1em}
\end{figure}

Fig. \ref{fig2} illustrates the convergence behaviors of the weighted sum rate via our proposed optimization method, with the transmit powers $P_U=5$ dBm and $P_B=10$ dBm, and different numbers of elements $L$ for the three schemes. 
All the curves grow fast, which validates the efficiency of the alternating optimization algorithm for maximizing the weighted sum rate.
In addition, since the IOSs bring more degrees of freedom, the cases with the DS-IOS and SS-IOS outperform those with WO-IOS, and evidently the weighted sum rate increases with $L$.
Besides, the rate with DS-IOS is significantly higher than that with SS-IOS, because all the elements of the DS-IOS reflect and refract signals from both sides simultaneously and independently, which fully exploits the reflecting and refracting resource of the intelligent surfaces.
\begin{figure}[t]
        \centering
        \includegraphics*[width=80mm]{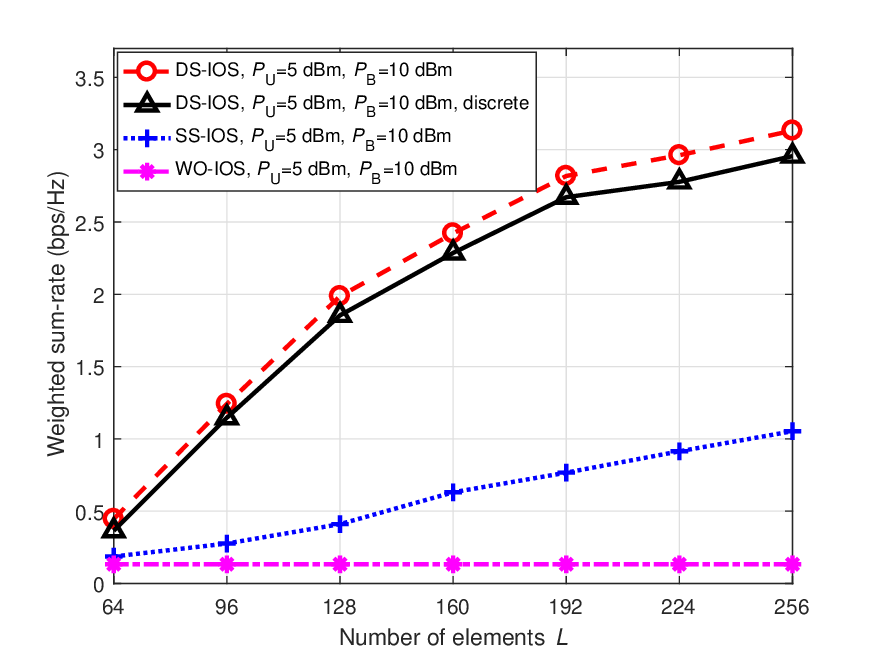}
       \caption{Three schemes with different $L$.}
        \label{fig3}
         \vspace{-1em}
\end{figure}

With different schemes and different numbers of elements $L$, the curves in Fig. \ref{fig3} show the performance of weighted sum rates.
From the curves, it is clear that data rates increase with the number of elements $L$. A significant improvement can be noticed via introducing the DS-IOS, compared to the traditional SS-IOS and the case without one. For example, the weighted sum rate can reach $3.13$ bps/Hz with $L=256$, $P_U=5$ and $P_B=10$ dBm of the DS-IOS, which is enhanced significantly compared to the case with an SS-IOS. 
Furthermore, taking into account the robustness of the proposed algorithm against discrete phase shifts,
we provide the comparison of continuous phase shifts and that with $4$ bit quantization. 
As we can see, a slight improvement is shown for the continuous case, compared to the discrete case with $4$ bit quantization using the same method as in \cite{twc}, for the DS-IOS case, e.g. $3.13$ bps/Hz compared to $2.95$ bps/Hz.

\begin{figure}[t]
        \centering
        \includegraphics*[width=80mm]{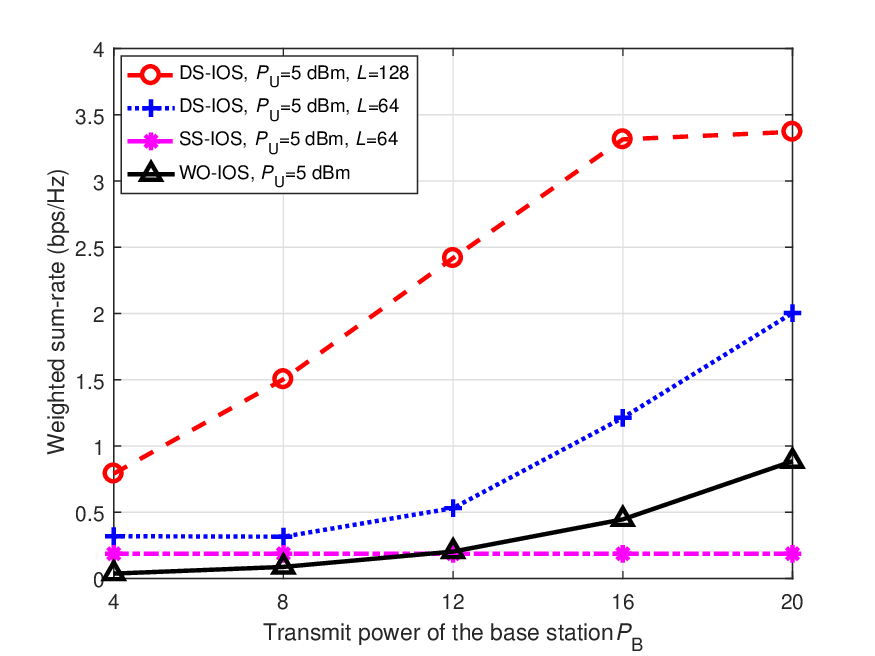}
       \caption{Three schemes with different downlink transmit power $P_B$.}
        \label{fig4}
         \vspace{-1em}
\end{figure}

\begin{figure}[t]
        \centering
        \includegraphics*[width=80mm]{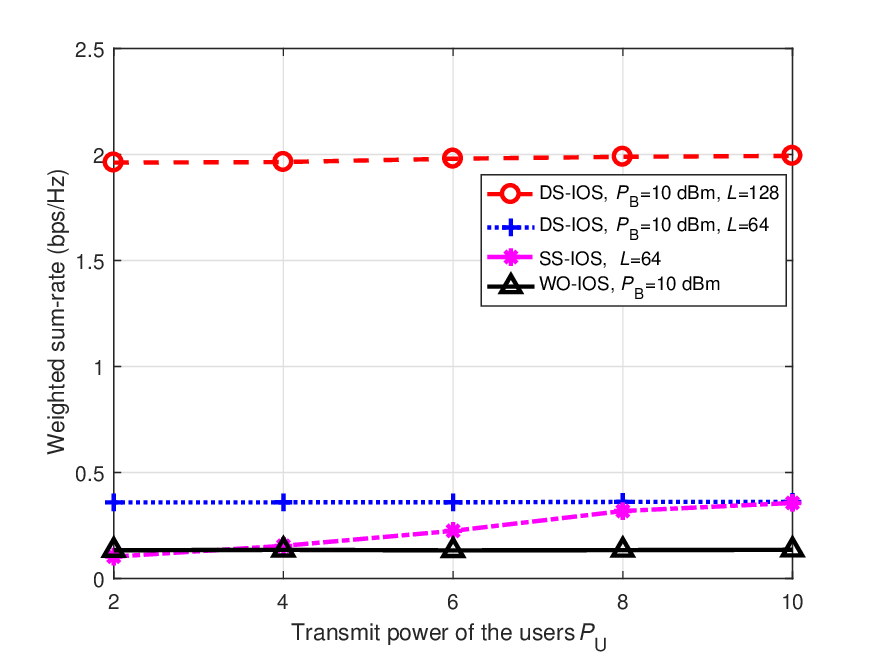}
       \caption{Three schemes with different uplink transmit power $P_U$.}
        \label{fig5}
         \vspace{-1em}
\end{figure}

In Fig. \ref{fig4}, the weighted sum rate performance of the proposed algorithm with $L=64$ and $128$ and different downlink transmit power $P_B$ for the three schemes is provided.
It is shown that the rate increases fast as the downlink transmit power grows. Furthermore, the SS-IOS focuses on uplink transmission because signals from only one side of it can be reflected and refracted; thus, $P_B$ would not impact the performance of SS-IOS. In addition, for $P_B<12$ dBm, the case without an IOS outperforms the case with an SS-IOS  because the SS-IOS only allows the uplink transmission. However, it is evident that our proposed DS-IOS achieves higher rates compared to the SS-IOS and WO-IOS. 
Furthermore, Fig. \ref{fig5} illustrates the relationship between the uplink transmit power $P_U$ and the weighted sum rate. 
We notice the rates with DS-IOS and WO-IOS remain constant as the uplink transmit power changes because when the downlink transmit power is larger than the uplink counterpart, the system utilizes more resources for downlink transmission. Moreover, the interference at the transmitter is more severe than that at the users.
\begin{figure}[t]
        \centering
        \includegraphics*[width=80mm]{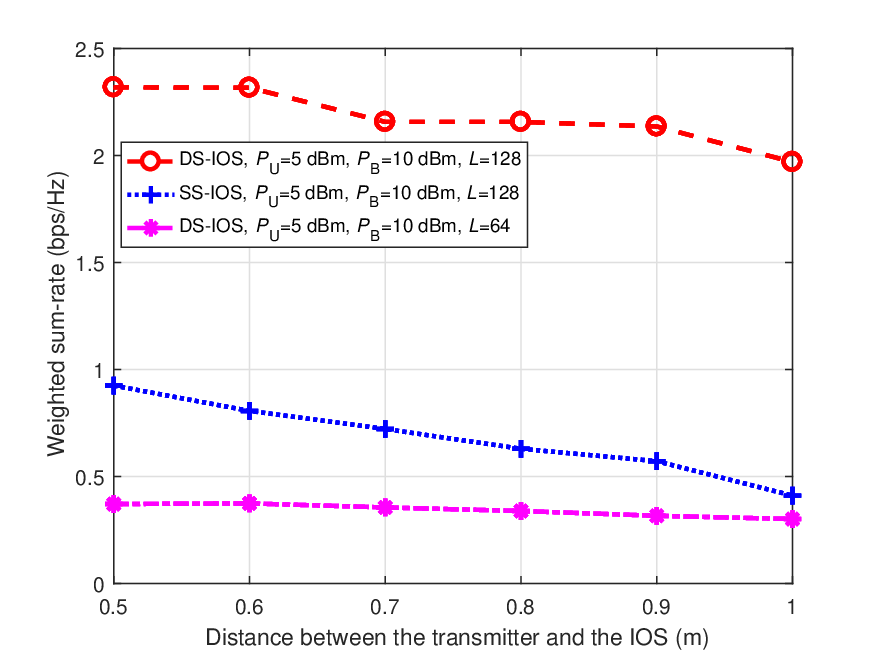}
       \caption{Three schemes with different distances between the transmitter and the IOS.}
        \label{fig6}
         \vspace{-1em}
\end{figure}
The effect of the distance between the transmitter and IOSs on the weighted sum rates is shown in the curves, as illustrated in Fig.  \ref{fig6}, which provide a critical guideline for designing the DS-IOS-assisted FD transmitter. Besides, comparing the case with DS-IOS and $L=128$ to that with SS-IOS and $L=128$, e.g., the system can obtain $2.16$ bps/Hz when the distance between the transmitter and the IOS is set as $0.8$ m. In contrast, the case with SS-IOS can only achieve $0.63$ bps/Hz, which implies that the proposed DS-IOS-assisted system significantly mitigates the SI problem in FD systems.

\section{Conclusion}
In this paper, a novel dual-side IOS assisted FD transmitter was proposed for a multi-user MIMO network, to solve the SWaP constraints and limited ability to cancel SI of the conventional FD schemes by completely bypassing the analogue SIC.
The weighted sum rate of the downlink and uplink transmissions of the FD system was maximized,
beamforming matrices at the transmitter and multiple users, as well as the components of the dual-side IOS were optimized.
However, the non-convexity and intractability due to the coupling of the variables bring some challenges.
Therefore, an iterative optimization algorithm was developed to tackle and solve the problem alternatively.
In specific, the Lagrangian dual approach was proposed to reduce computational complexity for optimizing the beamforming matrices.
Besides, amplitude and phase shift optimizations for reflection and refraction were still intractable;
hence, by exploiting the matrix property and convexifying the optimization problem, we solved it using the QCQP method.  
The effectiveness of our algorithm was validated by the results presented, demonstrating that the implementation of dual-side IOSs  significantly enhances the weighted sum-rate when compared to scenarios employing either SS-IOS or WO-IOS. This finding suggests a viable approach for improving spectrum efficiency in forthcoming IoT networks.

\begin{appendices}
\section{Derivations of \eqref{V_kd} and \eqref{V_ku}}
The objective function with respect to $\mathbf{V}_{k^d}$ and $\mathbf{V}_{k^u}$ can be rewritten as follows:
\begin{equation}\small\label{deriv1}\notag
\begin{gathered}
 f\left( {\left. {{{\mathbf{V}}_{{k^d}}},{{\mathbf{V}}_{{k^u}}}} \right|\forall k} \right) \hfill \\
   = \sum\limits_{k = 1}^K {{\gamma _{{k^d}}}\!{\text{Tr}}\left( {{{\mathbf{W}}_{{k^d}}}{\mathbf{V}}_{{k^d}}^H\!{\mathbf{H}}_{{k^d}}^H\!{{\mathbf{U}}_{{k^d}}}} \right)} \! +\! \sum\limits_{k = 1}^K {{\gamma _{{k^d}}}{\text{Tr}}\left( \!{{{\mathbf{W}}_{{k^d}}}\!{\mathbf{U}}_{{k^d}}^H\!{{\mathbf{H}}_{{k^d}}}\!{{\mathbf{V}}_{{k^d}}}} \!\right)}  \hfill \\
   \quad + \! \sum\limits_{k = 1}^K {{\gamma _{{k^u}}}{\text{Tr}}\left(\! {{{\mathbf{W}}_{{k^u}}}\!{\mathbf{U}}_{{k^u}}^H\!{{\mathbf{H}}_{{k^u}}}\!{{\mathbf{V}}_{{k^u}}}}\! \right)}  \!+\! \sum\limits_{k = 1}^K\! {{\gamma _{{k^u}}}\!{\text{Tr}}\left( \!{{{\mathbf{W}}_{{k^u}}}\!{\mathbf{V}}_{{k^u}}^H\!{\mathbf{H}}_{{k^u}}^H\!{{\mathbf{U}}_{{k^u}}}}\! \right)}  \hfill \\
\end{gathered} 
\end{equation}
\begin{equation}\small \notag
\begin{gathered}
 \quad  - \sum\limits_{k = 1}^K {{\gamma _{{k^d}}}} {\text{Tr}}\left( {{{\mathbf{W}}_{{k^d}}}{\mathbf{U}}_{{k^d}}^H{{\mathbf{H}}_{{k^d}}}{{\mathbf{V}}_{{k^d}}}{\mathbf{V}}_{{k^d}}^H{\mathbf{H}}_{{k^d}}^H{{\mathbf{U}}_{{k^d}}}} \right) \hfill \\
  \quad - \sum\limits_{k = 1}^K {{\gamma _{{k^d}}}} \sum\limits_{j = 1}^K {{\text{Tr}}\left( {{{\mathbf{W}}_{{k^d}}}{\mathbf{U}}_{{k^d}}^H{{\mathbf{H}}_{jk}}{{\mathbf{V}}_{{j^u}}}{\mathbf{V}}_{{j^u}}^H{\mathbf{H}}_{jk}^H{{\mathbf{U}}_{{k^d}}}} \right)}  \hfill \\
 \quad  - \sum\limits_{k = 1}^K {{\gamma _{{k^u}}}} \sum\limits_{j = 1}^K {{\text{Tr}}\left( {{{\mathbf{W}}_{{k^u}}}{\mathbf{U}}_{{k^u}}^H{{\mathbf{H}}_t}{{\mathbf{V}}_{{j^d}}}{\mathbf{V}}_{{j^d}}^H{\mathbf{H}}_t^H{{\mathbf{U}}_{{k^u}}}} \right)}  \hfill \\
\quad   - \sum\limits_{k = 1}^K {{\gamma _{{k^u}}}} \sum\limits_{j = 1}^K {{\text{Tr}}\left( {{{\mathbf{W}}_{{k^u}}}{\mathbf{U}}_{{k^u}}^H{{\mathbf{H}}_{{j^u}}}{{\mathbf{V}}_{{j^u}}}{\mathbf{V}}_{{j^u}}^H{\mathbf{H}}_{{j^u}}^H{{\mathbf{U}}_{{k^u}}}} \right)}  + {R_c} \hfill \\
\end{gathered}
\end{equation}
\begin{equation}\small
\begin{gathered}
   =  - \sum\limits_{k = 1}^K {\left( \begin{gathered}
  {\gamma _{{k^d}}}{\text{Tr}}\left( {{{\mathbf{W}}_{{k^d}}}{\mathbf{U}}_{{k^d}}^H{{\mathbf{H}}_{{k^d}}}{{\mathbf{V}}_{{k^d}}}{\mathbf{V}}_{{k^d}}^H{\mathbf{H}}_{{k^d}}^H{{\mathbf{U}}_{{k^d}}}} \right) \hfill \\
  \underbrace { + {\gamma _{{k^u}}}\sum\limits_{j = 1}^K {{\text{Tr}}\left( {{{\mathbf{W}}_{{k^u}}}{\mathbf{U}}_{{k^u}}^H{{\mathbf{H}}_t}{{\mathbf{V}}_{{j^d}}}{\mathbf{V}}_{{j^d}}^H{\mathbf{H}}_t^H{{\mathbf{U}}_{{k^u}}}} \right)} }_{{f_1}\left( {{{\mathbf{V}}_{{k^d}}}} \right)} \hfill \\ 
\end{gathered}  \right)} \! +\! {R_c} \hfill \\
 \quad  - \!\sum\limits_{k = 1}^K {\left( \begin{gathered}
  {\gamma _{{k^d}}}\!\sum\limits_{j = 1}^K {{\text{Tr}}\left( {{{\mathbf{W}}_{{k^d}}}\!{\mathbf{U}}_{{k^d}}^H\!{{\mathbf{H}}_{jk}}\!{{\mathbf{V}}_{{j^u}}}\!{\mathbf{V}}_{{j^u}}^H\!{\mathbf{H}}_{jk}^H\!{{\mathbf{U}}_{{k^d}}}} \!\right)}  \hfill \\
  \underbrace { +\! {\gamma _{{k^u}}}\!\sum\limits_{k = 1}^K {{\text{Tr}}\left(\! {{{\mathbf{W}}_{{k^u}}}\!{\mathbf{U}}_{{k^u}}^H\!{{\mathbf{H}}_{{j^u}}}\!{{\mathbf{V}}_{{j^u}}}\!{\mathbf{V}}_{{j^u}}^H\!{\mathbf{H}}_{{j^u}}^H\!{{\mathbf{U}}_{{k^u}}}}\! \right)} }_{{f_2}\left( \!{{{\mathbf{V}}_{{k^u}}}}\! \right)} \hfill \\ 
\end{gathered}  \right)}  \hfill \\
 \quad  +\! \sum\limits_{k = 1}^K {{\gamma _{{k^d}}}}\! {\text{Tr}}\left(\! {{{\mathbf{W}}_{{k^d}}}\!{\mathbf{V}}_{{k^d}}^H\!{\mathbf{H}}_{{k^d}}^H\!{{\mathbf{U}}_{{k^d}}}}\! \right) \!+\! \sum\limits_{k = 1}^K \!{{\gamma _{{k^d}}}} \!{\text{Tr}}\left(\! {{{\mathbf{W}}_{{k^d}}}\!{\mathbf{U}}_{{k^d}}^H\!{{\mathbf{H}}_{{k^d}}}\!{{\mathbf{V}}_{{k^d}}}}\! \right) \hfill \\
 \quad  +\! \sum\limits_{k = 1}^K\! {{\gamma _{{k^u}}}} \!{\text{Tr}}\!\left(\! {{{\mathbf{W}}_{{k^u}}}\!{\mathbf{U}}_{{k^u}}^H\!{{\mathbf{H}}_{{k^u}}}\!{{\mathbf{V}}_{{k^u}}}}\! \right) \!+\! \sum\limits_{k = 1}^K\! {{\gamma _{{k^u}}}}\! {\text{Tr}}\!\left( \!{{{\mathbf{W}}_{{k^u}}}\!{\mathbf{V}}_{{k^u}}^H\!{\mathbf{H}}_{{k^u}}^H\!{{\mathbf{U}}_{{k^u}}}}\! \right), \hfill \\
   \hfill \\ 
\end{gathered} 
\end{equation}
by setting
\begin{equation}\small
\begin{gathered}
  \sum\limits_{k = 1}^K {{f_1}\left( {{{\mathbf{V}}_{{k^d}}}} \right)\! =\! } \sum\limits_{k = 1}^K {{\gamma _{{k^d}}}{\text{Tr}}\left( {{{\mathbf{W}}_{{k^d}}}{\mathbf{U}}_{{k^d}}^H{{\mathbf{H}}_{{k^d}}}{{\mathbf{V}}_{{k^d}}}{\mathbf{V}}_{{k^d}}^H{\mathbf{H}}_{{k^d}}^H{{\mathbf{U}}_{{k^d}}}} \right)}  \hfill \\
  \qquad \qquad \qquad\ + \! \sum\limits_{k = 1}^k {{\gamma _{{k^d}}}\sum\limits_{j = 1}^K {{\text{Tr}}} \left( \! {{{\mathbf{W}}_{{k^d}}}\! {\mathbf{U}}_{{k^d}}^H\! {{\mathbf{H}}_{jk}}{{\mathbf{V}}_{{j^u}}}\! {\mathbf{V}}_{{j^u}}^H\! {\mathbf{H}}_{jk}^H{{\mathbf{U}}_{{k^d}}}}\!  \right)} , \hfill \\ 
\end{gathered} 
\end{equation}
\begin{equation}\small
\begin{gathered}
  \sum\limits_{k = 1}^K {{f_2}\left( {{{\mathbf{V}}_{{k^u}}}} \right) \!=\! } \sum\limits_{k = 1}^K {{\gamma _{{k^u}}}\sum\limits_{j = 1}^K {{\text{Tr}}\left( {{{\mathbf{W}}_{{k^u}}}{\mathbf{U}}_{{k^u}}^H{{\mathbf{H}}_t}{{\mathbf{V}}_{{j^d}}}{\mathbf{V}}_{{j^d}}^H{\mathbf{H}}_t^H{{\mathbf{U}}_{{k^u}}}} \right)} }  \hfill \\
   \qquad \qquad \qquad\ +\! \sum\limits_{k = 1}^K {{\gamma _{{k^u}}}} \sum\limits_{j = 1}^K {{\text{Tr}}} \left( \!{{{\mathbf{W}}_{{k^u}}}\!{\mathbf{U}}_{{k^u}}^H\!{{\mathbf{H}}_{{j^u}}}{{\mathbf{V}}_{{j^u}}}\!{\mathbf{V}}_{{j^u}}^H\!{\mathbf{H}}_{{j^u}}^H\!{{\mathbf{U}}_{{k^u}}}}\! \right), \hfill \\ 
\end{gathered} 
\end{equation}
and exploiting matrix properties, we have ${f_1}\left( {{{\mathbf{V}}_{{k^d}}}} \right) $ and ${f_2}\left( {{{\mathbf{V}}_{{k^u}}}} \right)$ in \eqref{deriv1} as follows:
\begin{equation}\small
\begin{gathered}
  {f_1}\left( {{{\mathbf{V}}_{{k^d}}}} \right) = {\text{Tr}}\left( {{\mathbf{V}}_{{k^d}}^H{\gamma _{{k^d}}}{\mathbf{H}}_{{k^d}}^H{{\mathbf{U}}_{{k^d}}}{{\mathbf{W}}_{{k^d}}}{\mathbf{U}}_{{k^d}}^H{{\mathbf{H}}_{{k^d}}}{{\mathbf{V}}_{{k^d}}}} \right) \hfill \\
 \qquad\qquad\quad  + {\text{Tr}}\left( {{\mathbf{V}}_{{k^d}}^H{\mathbf{H}}_t^H\left( {\sum\limits_{j = 1}^K {{\gamma _{{j^u}}}{{\mathbf{U}}_{{j^u}}}{{\mathbf{W}}_{{j^u}}}{\mathbf{U}}_{{j^u}}^H} } \right){{\mathbf{H}}_t}{{\mathbf{V}}_{{k^d}}}} \right), \hfill \\ 
\end{gathered} 
\end{equation}
\begin{equation}\small
\begin{gathered}
  {f_2}\left( {{{\mathbf{V}}_{{k^u}}}} \right) = {\text{Tr}}\left( {{\mathbf{V}}_{{k^u}}^H\left( {\sum\limits_{j = 1}^K {{\gamma _{{j^d}}}{\mathbf{H}}_{kj}^H{{\mathbf{U}}_{{j^d}}}{{\mathbf{W}}_{{j^d}}}{\mathbf{U}}_{{j^d}}^H{{\mathbf{H}}_{kj}}} } \right){{\mathbf{V}}_{{k^u}}}} \right) \hfill \\
 \qquad\qquad\quad  + {\text{Tr}}\left( {{\mathbf{V}}_{{k^u}}^H{\mathbf{H}}_{{k^u}}^H\left( {\sum\limits_{j = 1}^K {{\gamma _{{j^u}}}{{\mathbf{U}}_{{j^u}}}{{\mathbf{W}}_{{j^u}}}{\mathbf{U}}_{{j^u}}^H} } \right){{\mathbf{H}}_{{k^u}}}{{\mathbf{V}}_{{k^u}}}} \right), \hfill \\ 
\end{gathered} 
\end{equation}
respectively, which complete the derivation.

\section{Derivations of \eqref{g_theta_phi}}
Firstly, by re-writing the weighted sum rate function in \eqref{R_w} we have the following equation:
\begin{equation}\small
\begin{gathered}
  {R_w} = \sum\limits_{k = 1}^K {{\gamma _{{k^d}}}} \left( {{\text{Tr}}\left(\! {{{\mathbf{W}}_{{k^d}}}{\mathbf{V}}_{{k^d}}^H{\mathbf{H}}_{{k^d}}^H{{\mathbf{U}}_{{k^d}}}} \!\right) \!+\! {\text{Tr}}\left(\! {{{\mathbf{W}}_{{k^d}}}{\mathbf{U}}_{{k^d}}^H{{\mathbf{H}}_{{k^d}}}{{\mathbf{V}}_{{k^d}}}} \right)} \!\right) \hfill \\
\qquad\quad + \sum\limits_{k = 1}^K\! {{\gamma _{{k^u}}}} \!\left(\! {{\text{Tr}}\!\left( \!{{{\mathbf{W}}_{{k^u}}}\!{\mathbf{U}}_{{k^u}}^H\!{{\mathbf{H}}_{{k^u}}}\!{{\mathbf{V}}_{{k^u}}}} \!\right)\! + \!{\text{Tr}}\left(\! {{{\mathbf{W}}_{{k^u}}}\!{\mathbf{V}}_{{k^u}}^H\!{\mathbf{H}}_{{k^u}}^H\!{{\mathbf{U}}_{{k^u}}}} \!\right)}\! \right) \hfill \\
\qquad\quad   + \sum\limits_{k = 1}^K {{\gamma _{{k^d}}}} \sum\limits_{j = 1}^K {{\text{Tr}}\left( {{{\mathbf{W}}_{{k^d}}}{\mathbf{U}}_{{k^d}}^H{{\mathbf{H}}_{jk}}{{\mathbf{V}}_{{j^u}}}{\mathbf{V}}_{{j^u}}^H{\mathbf{H}}_{jk}^H{{\mathbf{U}}_{{k^d}}}} \right)}  \hfill \\
\qquad\quad   - \sum\limits_{k = 1}^K {{\gamma _{{k^d}}}{\text{Tr}}\left( {{{\mathbf{W}}_{{k^d}}}{\mathbf{U}}_{{k^d}}^H{{\mathbf{H}}_{{k^d}}}{{\mathbf{V}}_{{k^d}}}{\mathbf{V}}_{{k^d}}^H{\mathbf{H}}_{{k^d}}^H{{\mathbf{U}}_{{k^d}}}} \right)}  \hfill \\
\qquad\quad   - \sum\limits_{k = 1}^K {{\gamma _{{k^u}}}\sum\limits_{j = 1}^K {{\text{Tr}}\left( {{{\mathbf{W}}_{{k^u}}}{\mathbf{U}}_{{k^u}}^H{{\mathbf{H}}_{{j^u}}}{{\mathbf{V}}_{{j^u}}}{\mathbf{V}}_{{j^u}}^H{\mathbf{H}}_{{j^u}}^H{{\mathbf{U}}_{{k^u}}}} \right)} }  \hfill \\
\qquad\quad   - \sum\limits_{k = 1}^K {{\gamma _{{k^u}}}} \sum\limits_{j = 1}^K {{\text{Tr}}\left( {{{\mathbf{W}}_{{k^u}}}{\mathbf{U}}_{{k^u}}^H{{\mathbf{H}}_t}{{\mathbf{V}}_{{j^d}}}{\mathbf{V}}_{{j^d}}^H{\mathbf{H}}_t^H{{\mathbf{U}}_{{k^u}}}} \right)}  + {R_c}. \hfill \\ 
\end{gathered} 
\end{equation}

Then by substituting the channel coefficients ${{\mathbf{H}}_{{k^d}}} = {\mathbf{H}}_{i{u_k}}^H{{\mathbf{\Phi }}_t}{{\mathbf{H}}_{ti}},$ ${{\mathbf{H}}_{{k^u}}} = {\mathbf{H}}_{ir}^H{{\mathbf{\Phi }}_u}{{\mathbf{H}}_{{u_k}i}}$, ${{\mathbf{H}}_{jk}} = {{\mathbf{H}}_{{u_j}{u_k}}} + {\mathbf{H}}_{i{u_k}}^H{{\mathbf{\Theta }}_u}{{\mathbf{H}}_{{u_j}i}},$ and ${{\mathbf{H}}_t} = {{\mathbf{H}}_{tr}} + {\mathbf{H}}_{ir}^H{{\mathbf{\Theta }}_t}{{\mathbf{H}}_{ti}}$ into ${g_1} \left( {{{\mathbf{\Phi}}_t}} \right)$, ${g_2}\left( {{{\mathbf{\Phi}}_u}} \right)$, ${g_3}\left( {{{\mathbf{\Theta}}_u}} \right)$ and $ {g_4}\left( {{{\mathbf{\Theta}}_t}} \right)$, respectively. We have the transformations as follows:
\begin{equation}\small
\begin{gathered}
  {g_1}\left( {{{\mathbf{\Phi }}_t}} \right) \hfill \\
   = \sum\limits_{k = 1}^K {{\gamma _{{k^d}}}} {\text{Tr}}\left( {{{\mathbf{W}}_{{k^d}}}{\mathbf{V}}_{{k^d}}^H{\mathbf{H}}_{ti}^H{\mathbf{\Phi }}_t^H{{\mathbf{H}}_{i{u_k}}}{{\mathbf{U}}_{{k^d}}}} \right) \hfill \\
  \quad + \sum\limits_{k = 1}^K {{\gamma _{{k^d}}}} {\text{Tr}}\left( {{{\mathbf{W}}_{{k^d}}}{\mathbf{U}}_{{k^d}}^H{\mathbf{H}}_{i{u_k}}^H{{\mathbf{\Phi }}_t}{{\mathbf{H}}_{ti}}{{\mathbf{V}}_{{k^d}}}} \right) \hfill \\
 \quad   - \sum\limits_{k = 1}^K {{\gamma _{{k^d}}}} {\text{Tr}}\left( {{{\mathbf{W}}_{{k^d}}}{\mathbf{U}}_{{k^d}}^H{\mathbf{H}}_{i{u_k}}^H{{\mathbf{\Phi }}_t}{{\mathbf{H}}_{ti}}{{\mathbf{V}}_{{k^d}}}{\mathbf{V}}_{{k^d}}^H{\mathbf{H}}_{ti}^H{\mathbf{\Phi }}_t^H{{\mathbf{H}}_{i{u_k}}}{{\mathbf{U}}_{{k^d}}}} \right) \hfill \\
   = \sum\limits_{k = 1}^K {{\text{Tr}}\left( {{\mathbf{\Phi }}_t^H{\gamma _{{k^d}}}{{\mathbf{H}}_{i{u_k}}}{{\mathbf{U}}_{{k^d}}}{{\mathbf{W}}_{{k^d}}}{\mathbf{V}}_{{k^d}}^H{\mathbf{H}}_{ti}^H} \right)}  \hfill \\
  \quad  + \sum\limits_{k = 1}^K {{\text{Tr}}\left( {{{\mathbf{\Phi }}_t}{\gamma _{{k^d}}}{{\mathbf{H}}_{ti}}{{\mathbf{V}}_{{k^d}}}{{\mathbf{W}}_{{k^d}}}{\mathbf{U}}_{{k^d}}^H{\mathbf{H}}_{i{u_k}}^H} \right)}  \hfill \\
   \quad - \sum\limits_{k = 1}^K {{\text{Tr}}\left( {{\mathbf{\Phi }}_t^H{\gamma _{{k^d}}}{{\mathbf{H}}_{i{u_k}}}{{\mathbf{U}}_{{k^d}}}{{\mathbf{W}}_{{k^d}}}{\mathbf{U}}_{{k^d}}^H{\mathbf{H}}_{i{u_k}}^H{{\mathbf{\Phi }}_t}{{\mathbf{H}}_{ti}}{{\mathbf{V}}_{{k^d}}}{\mathbf{V}}_{{k^d}}^H{\mathbf{H}}_{ti}^H} \right)} , \hfill \\
   \hfill \\ 
\end{gathered} 
\end{equation}
\begin{equation}\small
\begin{gathered}
  {g_2}\left( {{{\mathbf{\Phi }}_u}} \right) \hfill \\
   = \sum\limits_{k = 1}^K {{\gamma _{{k^u}}}} {\text{Tr}}\left(\! {{{\mathbf{W}}_{{k^u}}}\!{\mathbf{U}}_{{k^u}}^H\!{{\mathbf{H}}_{{k^u}}}\!{{\mathbf{V}}_{{k^u}}}} \!\right) \!+\! \sum\limits_{k = 1}^K \!{{\gamma _{{k^u}}}} {\text{Tr}}\left( {{{\mathbf{W}}_{{k^u}}}\!{\mathbf{V}}_{{k^u}}^H{\mathbf{H}}_{{k^u}}^H{{\mathbf{U}}_{{k^u}}}} \right) \hfill \\
  \quad  - \! \sum\limits_{k = 1}^K {\sum\limits_{j = 1}^K {{\gamma _{{k^u}}}{\text{Tr}}\left( {{{\mathbf{W}}_{{k^u}}}{\mathbf{U}}_{{k^u}}^H{{\mathbf{H}}_{{j^u}}}{{\mathbf{V}}_{{j^u}}}{\mathbf{V}}_{{j^u}}^H{\mathbf{H}}_{{j^u}}^H{{\mathbf{U}}_{{k^u}}}} \right)} }  \hfill \\
  \end{gathered}
\end{equation}
\begin{equation} 
\begin{gathered}
   = \sum\limits_{k = 1}^K {{\text{Tr}}\left( {{{\mathbf{\Phi }}_u}{\gamma _{{k^u}}}{{\mathbf{H}}_{{u_k}i}}{{\mathbf{V}}_{{k^u}}}{{\mathbf{W}}_{{k^u}}}{\mathbf{U}}_{{k^u}}^H{\mathbf{H}}_{ir}^H} \right)}  \hfill \\
  \quad  + \sum\limits_{k = 1}^K {{\text{Tr}}\left( {{\mathbf{\Phi }}_u^H{\gamma _{{k^u}}}{{\mathbf{H}}_{ir}}{{\mathbf{U}}_{{k^u}}}{{\mathbf{W}}_{{k^u}}}{\mathbf{V}}_{{k^u}}^H{\mathbf{H}}_{{u_k}i}^H} \right)}  \hfill \\
 \quad   - \!\sum\limits_{k = 1}^K\! {\sum\limits_{j \!=\! 1}^K {{\text{Tr}}\!\left( \!{{\mathbf{\Phi }}_u^H\!{\gamma _{{k^u}}}\!{{\mathbf{H}}_{ir}}\!{{\mathbf{U}}_{{k^u}}}\!{{\mathbf{W}}_{{k^u}}}\!{\mathbf{U}}_{{k^u}}^H\!{\mathbf{H}}_{ir}^H\!{{\mathbf{\Phi }}_u}{{\mathbf{H}}_{{u_j}i}}\!{{\mathbf{V}}_{{j^u}}}\!{\mathbf{V}}_{{j^u}}^H\!{\mathbf{H}}_{{u_j}i}^H}\! \right)} } , \hfill \\ 
\end{gathered} 
\end{equation}
\begin{equation}\small
\begin{gathered}
  {g_3}\left( {{{\mathbf{\Theta }}_u}} \right) \hfill \\
   = \sum\limits_{k = 1}^K {\sum\limits_{j \!=\! 1}^K {{\text{Tr}}\left( \!{{\mathbf{\Theta }}_u^H{\gamma _{{k^d}}}\!{{\mathbf{H}}_{i{u_k}}}\!{{\mathbf{U}}_{{k^d}}}\!{{\mathbf{W}}_{{k^d}}}\!{\mathbf{U}}_{{k^d}}^H\!{\mathbf{H}}_{i{u_k}}^H\!{{\mathbf{\Theta }}_u}\!{{\mathbf{H}}_{{u_j}i}}{{\mathbf{V}}_{{j^u}}}\!{\mathbf{V}}_{{j^u}}^H\!{\mathbf{H}}_{{u_j}i}^H}\! \right)} }  \hfill \\
   \quad  + \sum\limits_{k = 1}^K {\sum\limits_{j = 1}^K {{\text{Tr}}\left( {{\mathbf{\Theta }}_u^H{\gamma _{{k^d}}}{{\mathbf{H}}_{i{u_k}}}{{\mathbf{U}}_{{k^d}}}{{\mathbf{W}}_{{k^d}}}{\mathbf{U}}_{{k^d}}^H{{\mathbf{H}}_{{u_j}{u_k}}}{{\mathbf{V}}_{{j^u}}}{\mathbf{V}}_{{j^u}}^H{\mathbf{H}}_{{u_j}i}^H} \right)} }  \hfill \\
   \quad  + \sum\limits_{k = 1}^K {\sum\limits_{j = 1}^K {{\text{Tr}}\left( {{{\mathbf{\Theta }}_u}{\gamma _{{k^d}}}{{\mathbf{H}}_{{u_j}i}}{{\mathbf{V}}_{{j^u}}}{\mathbf{V}}_{{j^u}}^H{\mathbf{H}}_{{u_j}{u_k}}^H{{\mathbf{U}}_{{k^d}}}{{\mathbf{W}}_{{k^d}}}{\mathbf{U}}_{{k^d}}^H{\mathbf{H}}_{i{u_k}}^H} \right)} }  \hfill \\
   \quad  + \sum\limits_{k = 1}^K {\sum\limits_{j = 1}^K {{\text{Tr}}\left( {{\gamma _{{k^d}}}{{\mathbf{W}}_{{k^d}}}{\mathbf{U}}_{{k^d}}^H{{\mathbf{H}}_{{u_j}{u_k}}}{{\mathbf{V}}_{{j^u}}}{\mathbf{V}}_{{j^u}}^H{\mathbf{H}}_{{u_j}{u_k}}^H{{\mathbf{U}}_{{k^d}}}} \right)} } , \hfill \\ 
\end{gathered} 
\end{equation}
\begin{equation}\small
\begin{gathered}
   {g_4}\left( {{{\mathbf{\Theta }}_t}} \right) \hfill \\ =\sum\limits_{k = 1}^K {\sum\limits_{j = 1}^K {{\gamma _{{k^u}}}{\text{Tr}}\left( {{{\mathbf{W}}_{{k^u}}}{\mathbf{U}}_{{k^u}}^H{\mathbf{H}}_{ir}^H{{\mathbf{\Theta }}_t}{{\mathbf{H}}_{ti}}{{\mathbf{V}}_{{j^d}}}{\mathbf{V}}_{{j^d}}^H{\mathbf{H}}_{ti}^H{\mathbf{\Theta }}_t^H{{\mathbf{H}}_{ir}}{{\mathbf{U}}_{{k^u}}}} \right)} }  \hfill \\
  \quad     + \sum\limits_{k = 1}^K {\sum\limits_{j = 1}^K {{\gamma _{{k^u}}}{\text{Tr}}\left( {{{\mathbf{W}}_{{k^u}}}{\mathbf{U}}_{{k^u}}^H{{\mathbf{H}}_{tr}}{{\mathbf{V}}_{{j^d}}}{\mathbf{V}}_{{j^d}}^H{\mathbf{H}}_{ti}^H{\mathbf{\Theta }}_t^H{{\mathbf{H}}_{ir}}{{\mathbf{U}}_{{k^u}}}} \right)} }  \hfill \\
   \quad      + \sum\limits_{k = 1}^K {\sum\limits_{j = 1}^K {{\gamma _{{k^u}}}{\text{Tr}}\left( {{{\mathbf{W}}_{{k^u}}}{\mathbf{U}}_{{k^u}}^H{\mathbf{H}}_{ir}^H{{\mathbf{\Theta }}_t}{{\mathbf{H}}_{ti}}{{\mathbf{V}}_{{j^d}}}{\mathbf{V}}_{{j^d}}^H{\mathbf{H}}_{tr}^H{{\mathbf{U}}_{{k^u}}}} \right)} }  \hfill \\
   \quad     + \sum\limits_{k = 1}^K {\sum\limits_{j = 1}^K {{\gamma _{{k^u}}}{\text{Tr}}\left( {{{\mathbf{W}}_{{k^u}}}{\mathbf{U}}_{{k^u}}^H{{\mathbf{H}}_{tr}}{{\mathbf{V}}_{{j^d}}}{\mathbf{V}}_{{j^d}}^H{\mathbf{H}}_{tr}^H{{\mathbf{U}}_{{k^u}}}} \right)} }, \hfill \\
\end{gathered}
\end{equation}respectively. In addition, we define the following expressions:
\begin{small}
\begin{equation*}
{{\mathbf{A}}_k} = {\gamma _{{k^d}}}{{\mathbf{H}}_{i{u_k}}}{{\mathbf{U}}_{{k^d}}}{{\mathbf{W}}_{{k^d}}}{\mathbf{U}}_{{k^d}}^H{\mathbf{H}}_{i{u_k}}^H,
\quad {{\mathbf{B}}_k} = {{\mathbf{H}}_{ti}}{{\mathbf{V}}_{k^d}}{\mathbf{V}}_{k^d}^H{\mathbf{H}}_{ti}^H,
\end{equation*}\end{small}
\begin{small}\begin{equation*}
{{\mathbf{C}}_k} =   {\gamma _{k^d}}{{\mathbf{H}}_{ti}}{{\mathbf{W}}_{k^d}}{\mathbf{U}}_{k^d}^H{\mathbf{H}}_{i{u_k}}^H, \qquad {{\mathbf{D}}_j} = {{\mathbf{H}}_{{u_j}i}}{{\mathbf{V}}_{ju}}{\mathbf{V}}_{j^u}^H{\mathbf{H}}_{{u_j}i}^H,
\end{equation*}\end{small}
\begin{small}\begin{equation*}
{{\mathbf{F}}_{kj}} = -{\gamma _{{k^d}}} {{\mathbf{H}}_{{u_j}i}}{{\mathbf{V}}_{j^u}}{\mathbf{V}}_{j^u}^H{\mathbf{H}}_{{u_j}{u_k}}^H{{\mathbf{U}}_{k^d}}{{\mathbf{W}}_{k^d}}{\mathbf{U}}_{k^d}^H{\mathbf{H}}_{i{u_k}}^H,
\end{equation*}\end{small}
\begin{small}\begin{equation*}
 {{\mathbf{Z}}_k} =   {\gamma _{k^u}}{{\mathbf{H}}_{{u_k}i}}{{\mathbf{V}}_{k^u}}{{\mathbf{W}}_{k^u}}{\mathbf{U}}_{k^u}^H{\mathbf{H}}_{ir}^H,
\end{equation*}\end{small}
\begin{small}\begin{equation*}
\begin{gathered}
  R_{cg} = \sum\limits_{k = 1}^K {\sum\limits_{j = 1}^K {{\gamma _{{k^d}}}{\text{Tr}}\left( {{\mathbf{V}}_{{k^u}}^H{\mathbf{H}}_{{u_k}{u_j}}^H{{\mathbf{U}}_{{j^d}}}{{\mathbf{W}}_{{j^d}}}{\mathbf{U}}_{{j^d}}^H{{\mathbf{H}}_{{u_k}{u_j}}}{{\mathbf{V}}_{{k^u}}}} \right)} }  \hfill \\
   \qquad\quad  + \sum\limits_{k = 1}^K {\sum\limits_{j = 1}^K {{\text{Tr}}\left( {{\gamma _{{k^u}}}{{\mathbf{W}}_{{k^u}}}{\mathbf{U}}_{{k^u}}^H{{\mathbf{H}}_{tr}}{{\mathbf{V}}_{{j^d}}}{\mathbf{V}}_{{j^d}}^H{\mathbf{H}}_{tr}^H{{\mathbf{U}}_{{k^u}}}} \right)} }  + {R_c}. \hfill \\
\end{gathered}
\end{equation*}\end{small}
Then we can reform the objective function with respect to the reflecting and refracting amplitudes and phase shifts of the dual-side IOS as in \eqref{g_theta_phi}, which completes the derivations.

\end{appendices}

\footnotesize
\bibliographystyle{IEEEtran}
\bibliography{ref}
\end{document}